\documentclass[11pt,preprint,letterpaper]{aastex}

\usepackage{epsfig}             
\usepackage{graphicx,color}     

\newcommand{\res}{\footnotesize \mbox{res} \normalsize}
\newcommand{\obs}{\footnotesize \mbox{obs} \normalsize}

\usepackage[hang,centerlast]{subfigure}          

\usepackage{natbib}
\usepackage{bibentry}

\usepackage{latexsym}

\shorttitle{Planetary Migration \& Secular Coupling}
\shortauthors{Agnor \& Lin}
\received{2010 December 13}
\accepted{2011 September 17}

\begin{document}

\title{On the Migration of Jupiter and Saturn: Constraints from Linear
  Models of Secular Resonant Coupling with the Terrestrial Planets}  

\author{Craig B.~Agnor$^{1}$ and D.~N.~C.~Lin$^{2,3}$}

\affil{$^1$Astronomy Unit, School of Physics \& Astronomy, Queen Mary
  University of London, UK; \email{C.B.Agnor@qmul.ac.uk}} 

\affil{$^2$Department of Astronomy and Astrophysics, University of
  California Santa Cruz, CA, USA}

\affil{$^3$Kavli Institute of Astronomy and Astrophysics \& College of 
  Physics, Peking University, Beijing, China}

\begin{abstract} 
We examine how the late divergent migration of
Jupiter and Saturn may have perturbed the terrestrial planets.
Using a modified secular model we have identified six secular
resonances  between the $\nu_5$ frequency of Jupiter and Saturn and
the four apsidal eigenfrequencies of the terrestrial planets
($g_{1\mbox{--}4}$).    We derive analytic upper limits on the eccentricity
and orbital migration timescale of Jupiter and Saturn when these
resonances were encountered to avoid perturbing the eccentricities of
the terrestrial planets to values larger than the observed ones.
Because of the small amplitudes of the $j=2,3$ terrestrial eigenmodes
the $g_2-\nu_5$ and $g_3-\nu_5$ resonances provide the strongest
constraints on giant planet migration.  If Jupiter and Saturn migrated
with eccentricities comparable to their present day values, smooth
migration with exponential timescales characteristic of
planetesimal-driven migration ($\tau\sim5\mbox{--}10$Myr) would have perturbed
the eccentricities of the terrestrial planets to values greatly
exceeding the observed ones.  This excitation may be mitigated if the
eccentricity of Jupiter was small during the migration epoch,
migration was very rapid (e.g., $\tau\lesssim 0.5$Myr perhaps via
planet\mbox{--}planet scattering or instability-driven migration) or the
observed small eccentricity amplitudes of the $j=2,3$ terrestrial
modes result from low probability cancellation of several large
amplitude contributions.  Results of orbital integrations show that
very short migration timescales ($\tau<0.5$ Myr), characteristic of
instability-driven migration, may also perturb the terrestrial
planets' eccentricities by amounts comparable to their  observed
values.  We discuss the implications of these constraints for the
relative timing of terrestrial planet formation, giant planet
migration, and the origin of the so-called Late Heavy Bombardment of
the Moon $3.9\pm0.1$ Ga ago.  We suggest that the simplest way to
satisfy these dynamical constraints may be for the bulk of any giant
planet migration to be complete in the first 30-100Myr of solar system
history.

\end{abstract}

\keywords{celestial mechanics \mbox{--} planetary systems: formation
  \mbox{--} planets and satellites: dynamical evolution and stability \mbox{--}
  planet\mbox{--}disk interactions \mbox{--} planets and satellites:
  formation \mbox{--} solar system: formation }

\section{Introduction}
\label{sec:intro}

During the last two decades, observational discoveries both external
to and within the solar system have provided strong evidence that
giant planets experience large-scale orbital migration after their
formation.  Outside the solar system the peculiar small orbits of the
so-called `Hot Jupiters' and their unlikely in situ formation
suggest that these gas giants underwent large-scale, radial migration
to their observed short-period orbits \citep[e.g.,][]{Lin_etal_1996}.
At early times ($\sim1\mbox{--}10$Myr) this might be accomplished via
interactions with the progenitor gas disk
\citep[e.g.,][]{Papaloizou_etal_2007}.  At later times migration may be
driven by scattering and redistribution of a planetesimal disk
\citep[e.g.,][]{Fernandez_&_Ip_1984,Murray_etal_1998,Hahn_&_Malhotra_1999}.
In addition, multi-planet systems may suffer orbital instabilities in
which giant planets undergo mutual scattering events that may eject
planets from  the system altogether and leave the surviving ones with
large orbital eccentricities and inclinations.  Such stochastic
evolution can account for the  orbital characteristics of many
observed extrasolar planets
\citep[e.g.,][]{Rasio_&_Ford_1996,Weidenschilling_&_Marzari_1996,Lin_&_Ida_1997,Levison_etal_1998}.

Closer to home, the resonant structure of the Kuiper Belt testifies to
the large-scale radial redistribution of the solar system's giant
planets more generally.  The eccentric and  inclined orbit of Pluto,
and the cohort of Plutinos in Neptune's exterior 3:2 mean motion
resonance, can be naturally explained by the outward migration of
Neptune \citep{Malhotra_1993,Malhotra_1995,Levison_etal_2008}.  As
the number of solar system characteristics that can be attributed to
processes associated with planet migration continues to increase,
there is a growing consensus that large-scale planetary migration did
occur in the solar system and that many of its dynamical structures
were determined during, and represent artifacts of, this epoch
\citep[e.g., see reviews by][]{Levison_etal_2007_ppv,Chiang_etal_2007}.

Due to their large masses and angular momenta, the giant planets are
the dominant perturbers of the solar system.  Therefore the period of
giant planet migration was one in which enormous orbital perturbations
swept through the solar system.  Clearly the timing and duration of
this epoch informs the evolution of the rest of the system.  

Because of their hydrogen and helium rich compositions, the formation
of Jupiter and Saturn must have been completed during the first few
million years of the solar system's history  while the gaseous
protoplanetary disk still persisted  \citep[e.g., see the review
by][]{Lissauer_&_Stevenson_2007}.  The accumulation of Uranus and
Neptune, is considerably less constrained \citep{Lissauer_etal_1995}
and remains an area of active research
\citep[e.g.,][]{Brydenetal_2000b,Ward_etal_2001,Thommes_etal_2003,Goldreich_etal_2004apj,Levison_etal_2007_ig}.
Estimates of the formation timescale of these planets reflect our
modest knowledge of their accumulation and span a wide range from
$\mathcal{O}(10^6\mbox{--}10^9)$ years
\citep{Goldreich_etal_2004apj,Lissauer_etal_1995}.   

Virtually all models suggested for the accretion of Uranus and Neptune
are inefficient at incorporating the mass of an initial planetesimal
disk into the ice giants on reasonably short timescales
(e.g.,~$\lesssim$10 Myr) and the scattering of these planetesimals may
lead to radial spreading and migration of the growing planets
\citep{Fernandez_&_Ip_1984}.  As a result, disk masses,  several times
that of the observed planets \citep[i.e.,~the so-called minimum mass
nebula;][]{Weidenschilling_1977} and formation locations  well
interior to their current orbits are often invoked to account for the
accretion of Uranus and Neptune.

However, local concentrations of planet building blocks at planet
trapping radii (such as the ice line or the interface between outward
and inward migration) may decrease the formation timescale of cores
with sufficient mass ($\gtrsim 10 M_\oplus$) to initiate rapid gas
accretion \citep{Kretke_2007,Lyra_2010}.  The formation of Jupiter
leads to gap formation in the solar nebula. Near the outer edge of
this gap, surface density and pressure maxima (at radii near or beyond
Jupiter's 2:1 MMR) also facilitate the accumulation of grains,
protoplanetary embryos and the formation of additional giant planet
cores \citep{Brydenetal_2000a,Brydenetal_2000b}.  This sequential
formation process may provide a scenario for the formation of the gas
and ice giant planets before the depletion of the gaseous nebula, with
modest, rather than extensive migration, mostly driven by their tidal
interaction with both residual disk gas and  planetesimals
\citep{Ida_etal_2000}.

The epoch, mode, and characteristic timescale in which the solar
system's giant planets migrated has important implications for the
evolution of the planetary system.  Giant planet migration models in
particle disks can be loosely divided into two classes, based on the
principal mode of migration.  Planetesimal-driven migration results
from the efficient scattering of small bodies by giant planets and is
characterized by the smooth radial divergence of planets embedded in a
planetesimal disk.  Recent models of this process in the solar system
suggest that the giant planets' orbits diverge with approximately
exponential migration timescales of $\tau\sim a/\dot{a}\simeq 5\mbox{--}20$Myr
\citep[e.g.,][]{Hahn_&_Malhotra_1999,Ida_etal_2000,Gomes_etal_2004}. 

In contrast to smooth radial planetesimal-driven migration, a global
instability among the giant planets, involving close encounters and
gravitational scattering between planets, results in large stochastic
jumps in the semimajor axes, eccentricities, and inclinations of the
planets
\citep{Thommes_etal_1999,Thommes_etal_2003,Tsiganis_etal_2005,Thommes_etal_2008_MMR2LHB,Batygin_&_Brown_2010}.
Because the planets traverse large changes in semimajor axis very
quickly, the bulk of planetary migration may be complete in a few
million years.

A recent effort to construct a comprehensive model of planetary
migration has succeeded in explaining several distinct dynamical
characteristics of the solar system as a product of specific giant
planet migration histories.  The so-called Nice model suggests that
the orbits of the giant planets evolved through phases of both
planetesimal-driven migration and planet\mbox{--}planet scattering
\citep[see][for a review]{Levison_etal_2007_ppv}.  In initial versions
of this model, the instability of the giant planets is initiated by
the divergent migration of Jupiter and Saturn across their mutual 2:1
mean motion resonance \citep{Tsiganis_etal_2005}. Dynamics associated
with the crossing of the 2:1 mean motion resonance (hereafter MMR)
play a fundamental role in the Nice model as  they are also
responsible for facilitating the capture of Jupiter's Trojan asteroids
with large orbital inclinations \citep{Morbidelli_etal_2005}.

In the inner solar system, planetary formation may have been completed
prior to the time when the giant planets reached their observed
orbits.  Both dynamical models of terrestrial planet accretion
\citep[e.g.,][]{Chambers_&_Wetherill_1998,Agnor_etal_1999,Chambers_2001,Raymond_etal_2004,Obrien_etal_2006,Hansen_2009,Raymond_etal_2009,Morishima_etal_2010}
and cosmochemical evidence
\citep[e.g.,][]{Kleine_etal_2002,Yin_etal_2002} suggest that
terrestrial planet formation was completed in $\sim30\mbox{--}100$Myr.

Uncertainties in the formation timescale of Uranus and Neptune
(e.g.~ranging from $10^6\mbox{ to }10^9$ yr) combined with estimates of
migration timescales ($\tau \sim 5-20$ Myr for planetesimal-driven
migration or much shorter for planet\mbox{--}planet scattering) suggest that
migration of the solar system's giant planets may have taken place
before, during, or after terrestrial planet formation.  Identifying
when and how giant planet migration occurred remains an outstanding
problem.

Indeed, the late orbital evolution of the solar system's giant planets
has even been invoked to account for cratering events on the Moon, an
explanation that necessarily requires fully formed terrestrial planets
to bear witness to the epoch of giant planet migration
\citep[e.g.,][]{Wetherill_1975}.  The so-called Late Heavy
Bombardment of the Moon (hereafter LHB) refers to a period from
$\sim3.9\pm0.1$ Ga ago when the lunar basins with known ages formed
\citep[\citeauthor{Tera_etal_1974} \citeyear{Tera_etal_1974}; see
also][for a detailed review]{Hartmann_etal_2000}.  The formation of
these basins may be attributed to a sharp increase in the cratering
rate of the Moon several hundred million years after planet formation
was complete.  The late orbital evolution of the giant planets has
been identified with this dramatic increase in the lunar cratering
rate via a variety of dynamical mechanisms including the sweeping of
resonances through the asteroid belt due to planetary migration
\citep{Strom_etal_2005,Minton_&_Malhotra_2009}, the late formation  of
Uranus and Neptune \citep{Levison_etal_2001}, orbital instabilities
among the giant planets resulting in orbit crossing, and planet\mbox{--}planet
gravitational scattering \citep{Thommes_etal_1999,Thommes_etal_2002},
or an orbital instability resulting from the divergent migration of
Jupiter and Saturn across their mutual 2:1 MMR as in the Nice model
\citep{Gomes_etal_2005}.  In each of these models, objects on
previously stable orbits are dislodged by the evolution of the giant
planets and delivered to Earth crossing orbits.  Implicit to each
model is the assumption that the terrestrial planets are unaffected by
the gross instability and/or changes in the giant planets' orbits
which are responsible for destabilizing the LHB impacting population.   
 
Here we examine this assumption in detail.  Specifically, we examine
how the divergent migration of Jupiter and Saturn from a more compact
configuration to their current orbits may perturb the terrestrial
planets. If the terrestrial planets witnessed an era of giant planet
migration, their present-day orbits offer two main dynamical
constraints: (1) the terrestrial planets must remain stable during the
epoch of migration and (2) the dynamical structure the terrestrial
planets emerging from the migration epoch must be consistent with the
observed terrestrial system.   Since the solar system's giant planets
migrate as a coupled system, any constraints on the migration of
Jupiter and Saturn may also be considered as implicit constraints on
the formation and migration of Uranus and Neptune. 

We develop linear secular models to identify migration histories of
Jupiter and Saturn that strongly couple them to the terrestrial
planets via apsidal secular resonances.   We adapt models for secular
resonant sweeping to predict the excitation of the terrestrial planets
as a function of the orbital state of Jupiter and Saturn at the time
of resonance and their migration timescale. Finally, we use these
models to identify migration histories that are consistent with the
observed orbital properties of the terrestrial planets.
 
This paper is organized in the following fashion.  In Section
\ref{sec:lin_theory}, we describe the secular model and its utility
for examining planetary migration.  In Section \ref{sec:secres}, we
discuss how passage through linear secular resonances excites
eccentricities and how these events may be used to constrain giant
planet migration histories.  In Sections \ref{sec:JS12} and
\ref{sec:g21}, we compare the predictions of the linear theory with
the results of $N$-body simulations.  In Section \ref{sec:disc}, we
evaluate model assumptions, discuss the implications of our prinicipal
results, and summarize our findings.  

Initial results, including the development of our secular model, the
identification of secular resonances, and some numerical simulations
were presented in a conference talk \citep{Agnor_&_Lin_2007}.
Subsequently, \citet{Brasser_etal_2009} have also examined this
problem within the context of the Nice model using $N$-body simulations.
Our linear model predictions are in general agreement with, and
provide a theoretical framework that accounts for, the results of
their simulation results. We compare the two where appropriate.

\section{Secular Evolution of the Planets}
\label{sec:lin_theory}

\subsection{Laplace\mbox{--}Lagrange Apsidal Precession}
The apsidal orbital precession of a system of planets can be described
using a set of linear first-order differential equations, whose
solution is a system of eigenmodes \citep[see,
e.g.,][]{Brouwer_&_Clemence_1961,Murray_&_Dermott_1999}.  
The equations of motion for the evolution of the planets'
eccentricities ($e$) and longitude of perihelia ($\varpi$) can be
constructed from the disturbing function 
by assuming that terms involving the mean longitudes are small and
retaining lowest-order terms in the eccentricity and inclinations.
For planet $i$, Lagrange's equations for the eccentricity vector $(h_i,k_i) =
     e_i(\sin\varpi_i, \cos\varpi_i)$ are
\begin{equation}
\frac{dh_{i}}{dt}  = \sum_{l} A_{il}k_{l}
\end{equation}
\begin{equation}
\frac{dk_{i}}{dt}  = - \sum_{l} A_{il}h_{l},
\end{equation}
where
\begin{equation}
A_{ii}= \frac{n_{i}a_{i}}{4} \sum_{l\ne i}
\left( \frac{m_l}{M_{\odot}+m_i} \right)
\frac{\alpha}{ a_{>} }  b_{3/2}^{1} 
      + \frac{3GM_{\odot}}{c^2a_{i}} n_i
\label{eq:akk}
\end{equation}
\begin{equation}
A_{il} = - \frac{n_i a_i}{4} 
           \left( \frac{m_l}{M_{\odot} + m_i}  \right)
                  \frac{\alpha}{  a_{>} }  b_{3/2}^{2},
\label{eq:aik}
\end{equation} 
and $m_i$, $a_i$, and $n_i$ are a planet's mass,
semimajor axis, and mean motion, respectively.    The standard terms
where $\alpha=a_</a_>$,  $a_{>}=$max$(a_{i},a_{l})$,
$a_{<}=$min$(a_{i},a_{l})$, and Laplace coefficients ($b^s_j$) have
been used.  Post-Newtonian corrections for relativistic precession are
included in the diagonal matrix elements where $c$ is the speed of
light \citep[see, e.g.,][]{Adams_&_Laughlin_2006}.

Solving these equations for a planetary system is an
eigenvalue problem whose eigenfrequencies ($g_{j}$) are real and represent the
precession frequencies of the modes. The solutions are of the form
\begin{equation}
h_{i} = \sum_{j} e_{ij}\sin(g_{j}t + \beta_{j})
\end{equation}
\begin{equation}
k_{i} = \sum_{j} e_{ij}\cos(g_{j}t + \beta_{j}),
\end{equation}
where the values of modal eccentricity amplitudes $e_{ij}$ and the
initial phases of the modes $\beta_{j}$ are constants determined
by projecting the system onto some initial state at time $(t=0)$.

\subsection{Eigenmode Amplitudes and the Angular Momentum Deficit}

In our use of this secular model there are two convenient and
complementary formulations of the eigenvalue solution, each with its
own advantages.  The modal eccentricity amplitude $(e_{ij})$ solution
(above) clearly identifies the orbital variation of each planet $i$
resulting from the coupling with the other planets via mode $j$.  As
the orbital elements of a system are its principal dynamical
coordinates, this  formulation is easily interpreted and can be
compared with the results of $N$-body simulations  (e.g.,~using Fourier
analysis to identify fundamental frequencies and amplitudes).
However, this representation of the system's modes tends to mask the
structural signature of the eigenvectors and the physical
interpretation of a mode's amplitude.

When considering the secular evolution of an isolated system the orbital
energies and semimajor axes of the planets are constant and the system
angular momentum is conserved.  These two constraints may be combined
to identify the so-called angular momentum deficit (AMD) as an important
integral of a secularly evolving system.  The AMD is defined as 
\begin{equation}
\mbox{AMD} = \sum_i m_i n_i a_i^2(1-\sqrt{1-e_i^2}\cos I_i),
\end{equation}
where $i$ runs over the planets and $I_i$ is a planet's orbital
inclination \citep[see, e.g.,][]{Laskar_1996}.
Physically, the AMD represents the angular momentum that must be added
to the system to make all the planets' orbits circular and coplanar.
Because secular interactions do not alter the semimajor axes
of the planets, a system's AMD quantifies the amount of angular
momentum that is available for exchange between planets via
gravitational interactions and limits the amplitude of eccentricity
and inclination oscillations of the planets.    

The linear secular system can also be formulated with the following
transformation of variables and matrix elements:
\begin{eqnarray}
H_i = \sqrt{m_i n_i a_i^2} e_i\sin \varpi_i =  \psi_i h_i \\ 
K_i = \sqrt{m_i n_i a_i^2} e_i\cos \varpi_i =  \psi_i k_i,
\end{eqnarray}
and $A_{il}\rightarrow(\psi_i/\psi_l)A_{il}$, where the factor $\psi_i$
is the square root of a planet's circular orbital angular momentum
\citep[see, e.g.,][]{Brouwer_&_Clemence_1961}.   

The system's solution is again expressed as a superposition of the
modes 
\begin{eqnarray}
H_i \equiv \sum_j H_{ij} = \sum_j C_j v_{ij} \sin(g_jt + \beta_j)\\
K_i \equiv \sum_j K_{ij} = \sum_j C_j v_{ij} \cos(g_jt + \beta_j),
\end{eqnarray}
where $\mathbf{v}_j\equiv\{v_{ij}\}$ is the $j$th orthonormal
eigenvector and $C_j$ its modal amplitude computed from a set of
orbital elements and masses.   The system quantity
$\frac{1}{2}\sum_im_in_ia_i^2e_i^2 =
\frac{1}{2}(H_i^2 + K_i^2)^2=\frac{1}{2}\sum_jC_j^2$ corresponds to the
horizontal component of the system's AMD in our secular model.  The
secular modal amplitudes $\{C_j\}$ are integrals of the system and reveal
the fraction of the system AMD partitioned into a single
mode. Conversion back and forth between the two 
formulations is easily accomplished using 
\begin{equation}
  e_{ij} = \frac{C_j v_{ij}}{\psi_i}.
\label{eq:eij}
\end{equation}

As the mass distribution of a planetary system evolves (e.g.,~due to
planetary migration, stellar spin down, dispersal of the gaseous
nebula, etc.), the system's eigenfrequencies, eigenvectors, and
modal amplitudes become time-dependent.  However, if the dynamical
tuning of the system is slow with respect to the eigenfrequencies
$\{g_j\}$ and the system is not near a secular resonance
(i.e.,~$g_j-g_k$ is not small), the variation in mode amplitudes tends
to oscillate rapidly and average out  \citep[see][for a complete
development]{Ward_1981}.  For slow enough changes in the system, the
modal amplitudes $\{C_j\}$ can be considered as effectively constant.
We exploit this convenient aspect of this formulation when exploring
the consequences of secular resonant sweeping during planetary
migration.

\subsection{Linear Solution of the Present-day Solar System}

We have calculated the normal mode solution for the eccentricity
evolution of a six-planet system consisting of Jupiter and Saturn and
the terrestrial planets.  Uranus and Neptune have been omitted from
this treatment for simplicity as their influence on Jupiter and Saturn
is weak and not efficiently communicated to the inner solar system.

\begin{figure}[!h]
\begin{center}
\resizebox{0.45\textwidth}{!}{\includegraphics{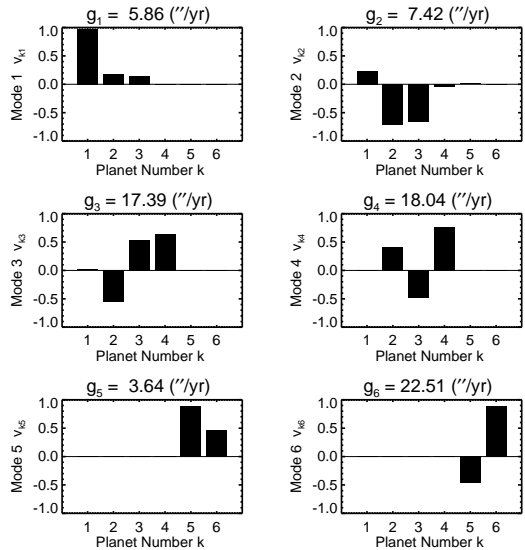}}  
\caption{Orthonormal eigenvectors of our six-planet model
  including Mercury through Saturn.  Each eigenvector is labeled with
  its eigenfrequency.  Large components for multiple
  planets indicate strong coupling between those
  planets in the mode.  The corrections due to nearby first-order mean
  motion resonances were included. \label{fig:e_vects}} 
\end{center}

\end{figure}

The system's orthonormal eigenvectors $(\mathbf{v}_j)$ are shown in
Figure \ref{fig:e_vects} with their associated eigenfrequencies
($g_j$).  As is the convention for the solar system, the eigenmode
index ($j$) identifies the planet that has the largest component in
the eigenvector (i.e.,~Mercury has the largest component in the $j=1$
mode and its evolution is strongly governed by this mode, Venus has
the largest component in the $j=2$ mode, etc.).  Large
components for multiple planets in a single eigenvector indicate
strong coupling between those planets.  The sign of an eigenvector
component ($v_{ij}$) indicates the relative apsidal orientation within
the mode, with like signs indicating alignment and opposite signs
anti-alignment.

The eigenvectors of this six-planet system can be divided into two
groups, those where the terrestrial planets have the largest
components ($j=1\mbox{--}4$) and those where Jupiter and Saturn have the
dominant components ($j=5\mbox{--}6$).  This trend in the structure  of the
eigenvectors is indicative of the weak coupling between the
terrestrial and giant planets in their current configuration.  The
eigenvector signatures of the terrestrial modes ($j=1\mbox{--}4$) each have
large components for at least two planets indicating significant
coupling in their evolution.

\begin{figure}[!h]
\begin{center}
\resizebox{0.45\textwidth}{!}{\includegraphics{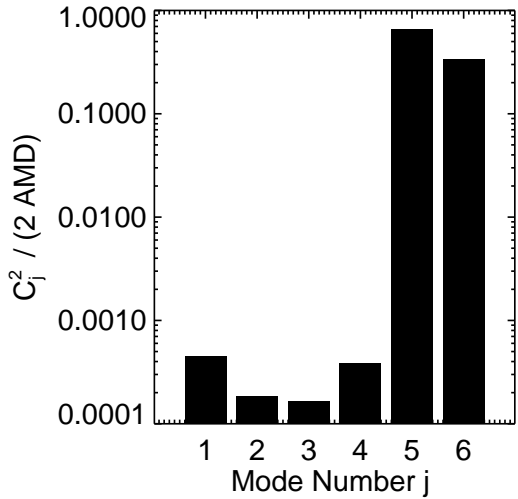}}         
\end{center}
\caption{Fraction of the solar system's angular momentum deficit
  (AMD) in each eccentricity mode is shown.  Note: $\mbox{AMD}=\sum_jC_j^2/2$. \label{fig:e_amps}} 
\end{figure}

Using the VSOP82 ephemeris of \citet{Bretagnon_1982}, we have computed
the initial amplitudes and phases of this system's normal modes.  The
fraction of the system's AMD present in
each mode ($C^2_j$/2) is shown in Figure \ref{fig:e_amps}.    The
modal eccentricity amplitudes are listed in Table \ref{tab:eij}.  The
$j=5,\,6$ modes of Jupiter and Saturn contain the majority of the
system's AMD.\footnote{The neglected  contributions of Uranus and
Neptune (and the resulting $j=7,\,8$ modes) amount to an additional 2\%
of the system AMD.}   Among the modes that control the evolution of
the terrestrial planets the $j=1$ and $j=4$ mode amplitudes contain
$\sim85$\% of the terrestrial subsystem's AMD (i.e.,~the sum
$\Sigma_{j=1}^4 C_j^2/2$).  The $j=1$ and $j=4$ modes strongly affect
the evolution of Mars and Mercury and are responsible for the large
eccentricities of these planets.  Similarly, the $j=2,\,3$ modes
strongly affect the evolution of Venus and Earth.  The relatively
small amplitudes of these modes are manifested as smaller
eccentricities of Venus and Earth.    

This dynamical configuration and partitioning of the AMD between the
terrestrial planets is peculiar.   The terrestrial eigenfrequencies
with the largest amplitudes are both in close proximity to an
eigenfrequency  with a much smaller amplitude.   For example, the
$j=1$ and $j=2$ frequencies are close with $g_1=5.86^{\arcsec} \mbox{yr}^{-1}$ and
$g_2=7.42^{\arcsec}\mbox{yr}^{-1}$, and the amplitude ratio is $C_1/C_2\approx5$.
Similarly, the $j=3$ and $j=4$ modes that strongly affect Earth and
Mars are close in value with $g_3=17.39^{\arcsec}\mbox{yr}^{-1}$ and
$g_4=18.04^{\arcsec}\mbox{yr}^{-1}$ and have an amplitude ratio of
$C_4/C_3\approx2$.   

\begin{table}[!ht]
\begin{center}
\begin{tabular}{|rc|rrrrrr|}
\hline
\multicolumn{2}{|c|}{ }& \multicolumn{6}{c|}{Mode Index $j$} \\

    Planet   &$i$  &    \multicolumn{1}{c}{1}  &  \multicolumn{1}{c}{2} & \multicolumn{1}{c}{3} &  \multicolumn{1}{c}{4} &
   \multicolumn{1}{c}{5} &        \multicolumn{1}{c|}{6}  \\
\hline 
    Mercury   & 1  &   0.1808  &   0.0216  &   0.0164  &   0.0015  &  --0.0373  &  --0.0162  \\
      Venus   & 2  &   0.0097  &  --0.0185  &  --0.0166  &  --0.0017  &   0.0191  &   0.0094  \\
      Earth   & 3  &   0.0005  &  --0.0119  &   0.0111  &   0.0202  &  --0.0017  &  --0.0066  \\
       Mars   & 4  &   0.0010  &  --0.0248  &   0.0277  &  --0.0673  &   0.0012  &   0.0100  \\
    Jupiter   & 5  &   0.0000  &   0.0000  &   0.0000  &   0.0000  &   0.0433  &   0.0162  \\
     Saturn   & 6  &   0.0000  &  --0.0000  &   0.0000  &   0.0000  &  --0.0357  &   0.0484  \\
\hline
\end{tabular}
\caption{Eccentricity Amplitudes ($e_{ij}$) Computed in Our Six-planet
  Secular Model Using the Ephemeris of
  \citet{Bretagnon_1982}  \label{tab:eij}} 
\end{center}
\end{table}

The frequencies of the terrestrial modes (i.e.,~$j=1\mbox{--}4$) all take
values \emph{between} those that dominate the present--day evolution of
Jupiter and Saturn  (i.e.,~$g_5 < g_{1\mbox{--}4} < g_6$).  As a result the
giant planet precession frequencies need not be grossly modified to
resonate with terrestrial modes (e.g.,~$g_5=g_1$).  Meteorites and
asteroids placed in or near a secular resonance with the giant planets
may be strongly perturbed to eccentric orbits
\citep{Farinella_etal_1993}, and may even be driven into the Sun
\citep[e.g.,][]{Gladman_etal_1997,Chambers_&_Wetherill_1998}.     

Linear secular theory has a number of properties that make it a useful
tool for exploring the implications of giant planet migration.  As
described above, the secular frequencies $\{g_j\}$ and eigenvectors
$\{\mathbf{v}_j\}$ are functions of the planet masses $\{m_i\}$ and
semimajor axes $\{a_i\}$.  This allows us to compute the evolution of
the system's eigenfrequencies, and their sweeping through the solar
system, for various orbital migration histories of the planets.  When
a system's mass distribution does not change, its mode amplitudes
$\{C_j\}$ are integrals of the system's time evolution.  For the solar
system, the planetary precession periods are $\sim10^4\mbox{--}10^5$yr.  As
long as the mass distribution of the system  evolves on longer
timescales and the system is not near resonance, the modal amplitudes
$\{C_j\}$  may be considered effectively  constant \citep{Ward_1981}.

\subsection{Secular Precession Near First-order Mean Motion Resonances}

In the standard Laplace--Lagrange model the direct terms of the
disturbing function containing the mean longitudes and all the
indirect terms in the disturbing function may be neglected when
averaging over long timescales.  However, near commensurabilities
between the mean motions of planets, resonant terms in the disturbing
function can make significant contributions to orbit precession.  A
well-known example of this in the solar system is the so-called Great
Inequality which results from the close proximity of Jupiter and
Saturn to their mutual 5:2 MMR.  Jupiter and Saturn's nearness to this
resonance modifies the $g_5$ and $g_6$ frequencies by
$\approx+0.5\arcsec$yr$^{-1}$ and $+6\arcsec$yr$^{-1}$ respectively \citep[see,
e.g.,][]{Brouwer_&_Clemence_1961,Murray_&_Dermott_1999}.     The
lowest-order  mean motion resonances encountered during the divergent
migration of Jupiter and Saturn are the 2:1 and 3:2.  Both are
first-order resonances and may strongly affect the secular evolution
of the planets.   

\citet{Malhotra_etal_1989} developed an approach for including the
influence  of first-order mean motion resonances in the secular model
and  used it to account for the evolution of the Uranian satellites.
Their treatment of the problem includes the averaged influence of the
direct terms from the disturbing function related to first-order
($\ell:\ell-1$) mean motion resonances. The 2:1 is unusual among the
first-order resonances as it also gives rise to indirect terms in the
disturbing function of both planets
\citep[cf.][]{Murray_&_Dermott_1999}.  

\begin{figure}[!h]
\begin{center}
\subfigure{
\resizebox{0.5\textwidth}{!}{\includegraphics{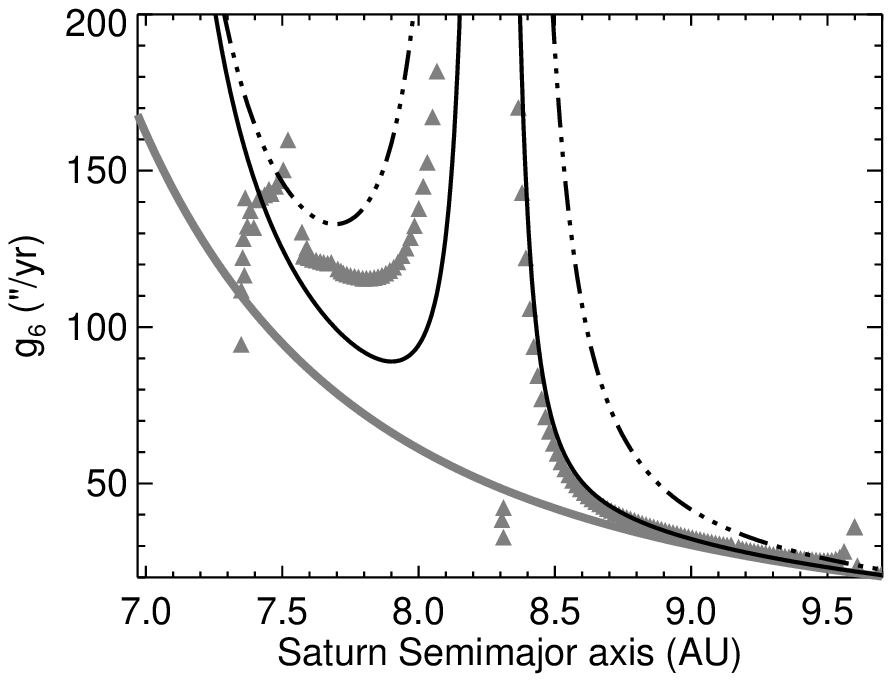}}}\\
\subfigure{
\resizebox{0.5\textwidth}{!}{\includegraphics{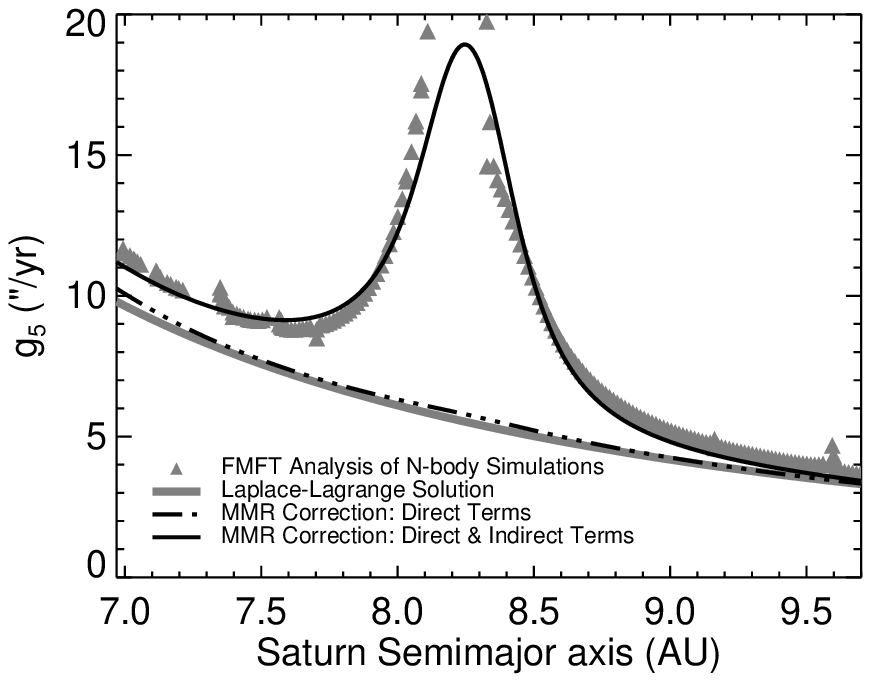}}}
\end{center}
\caption{Comparing results of modified secular theory and $N$-body
  simulations.  Jupiter orbits with a semimajor axis of 5.2 AU.
  Saturn's semimajor axis is changed from 7 to 9.7 AU.  The apsidal
  eigenfrequencies of the $g_6$ and $g_5$ modes were computed using
  standard Laplace--Lagrange theory (gray solid line), with the direct
  terms of from the disturbing function of first-order mean motion
  resonances (dash-dotted black line) and with both direct and indirect
  contributions from first-order mean motion resonances (black solid
  lines).  Also shown are results from individual $N$-body
  integrations that have been Fourier analyzed to identify the
  characteristic precession frequencies of Jupiter and Saturn (gray triangles). 
  \label{fig:fmftg56}} 
\end{figure}

Figure \ref{fig:fmftg56} shows the apsidal eigenfrequencies of Jupiter
and Saturn as a function of Saturn's semimajor axis determined using
several different methods.  In each case, Jupiter's semimajor axis was
held constant at 5.2 AU, the semimajor axis of  Saturn was increased
from just outside the mutual 3:2 mean motion resonance at about 6.8
AU, across the 2:1 at 8.25 AU, to 9.7 AU.  

To determine the eigenfrequencies of the Jupiter--Saturn system as a
function of their semimajor axes we performed a set of individual
$N$-body simulations, each with a different initial semimajor axis for
Saturn.  In each simulation the planets were assigned nearly circular,
nearly coplanar orbits (i.e.,~$e\simeq0.005$, $I=e/2$), random phase
elements and were  evolved for about 11 Myr.  The SyMBA integrator
\citep{Duncan_etal_1998} was used in these simulations and the
precessional effects of general relativity were included
\citep[e.g.,][]{Quinn_etal_1991}.   

We determined the eccentricity amplitudes ($e_{ij}$) and
eigenfrequencies of the system by analyzing the evolution of a
planet's eccentricity vector $(h_i,k_i)$ using a Frequency Modified
Fourier Transform code \citep[hereafter
FMFT;][]{Sidlichovsky_&_Nesvorny_1997}.\footnote{The FMFT code used in
this study was derived from that used in
\citet{Christou_&_Murray_1999}.}   For each simulation the two
frequencies with the largest amplitudes in the evolution of Jupiter's
eccentricity vector $(h_5,k_5)$ are shown in Figure
\ref{fig:fmftg56} with gray triangles.  The close proximity of the
measured frequencies to the eigenfrequencies of the Jupiter--Saturn
system identifies them as the $g_5$ and $g_6$.   Since the $N$-body
simulations include all gravitational interactions between the
planets, resonant, and otherwise, these FMFT results serve as
``measurements'' of the apsidal eigenfrequencies of the actual system
and can be used to assess the fidelity of more approximate, but
analytically tractable, treatments.   

As Saturn's orbit is expanded, the values of the $g_5$ and $g_6$
frequencies show two important characteristics.  First, near both the
3:2 and 2:1 MMRs  the value of the $g_6$ frequency diverges.  The
behavior of the $g_5$ is somewhat different, showing divergence
behavior only near the 2:1 mean motion resonance.  For orbital
separations outside the 2:1 resonance (i.e.,~the semimajor axis of
Saturn greater than 8.25AU in this case), the two eigenfrequencies
decrease as Jupiter and Saturn continue to diverge.  

We have used secular theory to compute the modal frequencies in three
different ways.  The standard Laplace--Lagrange treatment is shown as a
gray solid line in both figures.  This treatment contains no
information regarding mean longitudes of the planets and fails to
account for divergence in the $g_{5,6}$ frequencies near the 3:2 and
2:1 mean motion resonances.  

We also computed the apsidal eigenfrequencies of Jupiter and Saturn,
correcting the Laplace--Lagrange treatment with the direct terms of
\citet{Malhotra_etal_1989}.  The value of $g_5$ and $g_6$ are shown
with a dash-dotted line.  The corrections due to the direct terms
account for divergence of the $g_6$ frequency when approaching the 3:2
or 2:1 resonances, but fail to account for the changes in the $g_5$
frequency due to the 2:1 resonance.

Finally, we compute the apsidal eigenfrequencies using the correction
of both direct and indirect terms from the first-order mean motion
resonances.  The contributions of the indirect terms associated with
the 2:1 MMR may be included in the model of \citet{Malhotra_etal_1989}
by adding $-2\alpha\delta_{j,1}$ to their Equation~(26b) where $\delta_{j1}$
is the kronecker delta function (R.~Malhotra, private
communication 2011).\footnote{A more cumbersome inclusion of the indirect
terms that yields similar results may also be derived from the
disturbing function for osculating elements \citep[see,
e.g.,][]{Ellis_&_Murray_2000,Murray_&_Dermott_1999}.}  As the 2:1 is
the only first-order resonance with indirect terms it is also the only
resonance affected by this amended theory.  

The inclusion of the indirect terms dramatically improves the
agreement between the secular frequencies computed with FMFT-measured
values and the corrected secular theory and accounts for the behavior
of the $g_5$ frequency around the 2:1 MMR.  This model assumes that
the resonant argument is rapidly circulating and consequently becomes
less accurate near the resonant separatrix where the circulation rate
slows.  The continuous behavior of the corrected $g_5$ frequency as
the 2:1 MMR is crossed, its inflection and local maxima are model
artifacts of these assumptions.   For Jupiter and Saturn more than
$\sim0.1$ AU outside the 2:1 MMR resonance, the results of the $N$-body
simulations and the modified secular theory are in close agreement.  

The influence of additional mean motion resonances between Jupiter and
Saturn are also visible in FMFT  measurements of $g_5$ and $g_6$.
These appear as deviation from the smooth general trends indicated by
secular theory.  The effects of the 5:3 and 7:4 MMRs when Saturn is
near 7.3 and 7.55 AU and the 5:2 MMR as Saturn approaches its current
orbit at 9.54 AU are evident.  These second- and third-order resonances
are relatively weak compared to the 3:2 and 2:1 and their influence on
the eigenfrequencies is both smaller and localized to configurations
close to exact resonance. 

These results clearly demonstrate that the divergent migration of
Jupiter and Saturn across their 2:1 MMR, as suggested by
\citet{Tsiganis_etal_2005} and \citet{Gomes_etal_2005}, results in
secular resonant sweeping of the $g_5$ across the values of all four
of the current terrestrial eigenfrequencies.   Thus, this migration
scenario may subject the terrestrial planets to potentially
destabilizing secular resonant perturbations.

\section{Excitation Due to Secular Resonant Crossing}
\label{sec:secres}

Excitation via passage through linear secular resonances with the
giant planets has been considered previously in several contexts
\citep{Ward_etal_1976,Heppenheimer_1980,Nagasawa_etal_2005}.
First, we describe the test particle solutions of \citet{Ward_etal_1976} and 
\citet{Nagasawa_etal_2005} adapting them to the case of resonant
sweeping due to planetary migration.\footnote{See also \citet{Gomes_1997} and
  the recent work of \citet{Minton_&_Malhotra_2011} that have used similar
  models to evaluate the migration induced sweeping of the $g_6$ and $s_6$ resonances
  through the asteroid belt.}  We then extend this model
to describe resonances with the eigenmodes of the terrestrial planets
using the subsystem approximation of \citet{Ward_1981}.

\subsection{Test Particle Excitation}

Consider a test particle in a system where planetary migration causes
its orbital precession frequency ($g$) to pass through a resonance
with one of the system's eigenfrequencies ($g_{j}$).  In a reference
frame that rotates with frequency $g_{j}$, the equations of motion of
$h = e\sin(\varpi-g_{j}t-\beta_{j})$ and
$k=e\cos(\varpi-g_{j}t-\beta_{j})$ for the planetesimal interacting 
with the resonant mode are
\begin{equation}
\frac{dh}{dt}-(g-g_{j})k = \mu_{j}
\label{eq:dh1}
\end{equation}
\begin{equation}
\frac{dk}{dt}+(g-g_{j})h = 0,
\label{eq:dh2}
\end{equation}
where 
\begin{equation}
\mu_{j} 
        = \frac{n a}{4} \sum_i e_{ij} \left(
          \frac{m_i}{M_{\odot}}\right) \frac{\alpha}{a_{>}} b_{3/2}^2.
\label{eq:nuj}
\end{equation}

An analytic solution for resonance passage can be obtained by introducing a
new independent variable $\phi \equiv \int \Delta g dt$, where $\Delta
g = g - g_j$ and transforming Equations \ref{eq:dh1} and \ref{eq:dh2} into
\begin{equation}
\frac{dh}{d\phi} - k = \frac{\mu_{j}}{\Delta g}
\end{equation}
\begin{equation}
h  + \frac{dk}{d \phi}=0.
\end{equation}
The particular solution of this expression is
\begin{equation}
h = (C_1 + c)\cos \phi + (S_1 + s) \sin \phi
\end{equation}
\begin{equation}
k = -(C_1 + c)\sin\phi + (S_1 + s) \cos \phi
\end{equation}
with
\begin{equation}
 C_1 = \int \left(\frac{\mu_j}{\Delta g} \right) \cos \phi\, d\phi
\end{equation}
\begin{equation}
 S_1 = \int \left(\frac{\mu_j}{\Delta g} \right) \sin \phi\, d\phi.
\end{equation}
Defining $t=0$ as the time of exact resonance ($g=g_{j}$) and setting
$\Delta \dot{g} =\dot{g} -\dot{g}_j$ where $\dot{g}_j=|dg_j/dt|$, the variable  
$\phi=-t'^2=-\frac{1}{2}\Delta \dot{g}t^2=-\frac{1}{2}(\Delta
g)^2/\Delta \dot{g}$, the solutions for $h,k$ following resonance
passage are
\begin{equation}
h_{\res} = -\mu_{j} \left( \frac{\pi}{\Delta \dot{g}} \right)^{1/2} 
\left[  \left(  C_1(t') + \frac{1}{2}  \right)  \cos t'^2
      +  \left(  S_1(t') + \frac{1}{2}  \right) \sin t'^2 \right]
\end{equation}
\begin{equation}
k_{\res} = \mu_{j} \left( \frac{\pi}{\Delta \dot{g}} \right)^{1/2} 
\left[  \left(  S_1(t') + \frac{1}{2}  \right)  \cos t'^2
      -  \left(  C_1(t') + \frac{1}{2}  \right) \sin t'^2 \right]
\end{equation}
where $S_{1}$ and $C_{1}$ are the Fresnel sine and cosine integrals
and we have required that $h,k\rightarrow0$ as $t\rightarrow -\infty$.

Long after resonance passage, as $t\rightarrow \infty$, the
test particle's new free eccentricity is
\begin{equation}
 e_{\res} = \mu_{j} \left( \frac{2 \pi}{|\Delta \dot{g}|}
 \right)^{1/2}. 
 \label{eq:eres}
\end{equation}
In essence, resonance passage excites a new contribution to the free
eccentricity of the body.  The amplitude of this response is determined by the
rate of passage through resonance and is a function of the migration rate.   
It is convenient to use an exponentially decaying migration rate of the
form 
\begin{equation}
\frac{da_i}{dt}=\frac{\Delta a_i}{\tau},
\label{eq:adot_tau}
\end{equation}  
where $\Delta a_i = a_{i,f} - a_{i}$ is the distance
of the planet from its final semimajor axis ($a_{i,f}$) and $\tau$ is the
characteristic migration timescale.  We treat $\tau$ as a global
migration constant and use this to parameterize migration speed. We
note that Equation (\ref{eq:adot_tau}) can also be used to convert between
analytic constraints, that are easily expressed as an exponential
timescale ($\tau$), and migration rates ($da_i/dt$) which might be
more easily measured from $N$-body simulations of planetary migration.
Using the exponential migration model, 
\begin{equation}
  \Delta \dot{g} = \sum_i \left( \frac{dg}{da_i} -  \frac{dg_j}{da_i}
  \right) \frac{\Delta a_i}{\tau},
\end{equation}
where all terms except the timescale are functions of the
planetary masses and semimajor axes at the time of resonance. This can
also be used to identify a critical migration timescale ($\tau^{*}$)
required to excite the eccentricity to some prescribed value $e_{\res}$
\begin{equation}
\tau^{*} = \frac{1}{2\pi} \left|\sum_i \left(\frac{dg}{da_i} -
                  \frac{dg_j}{da_i} \right) \Delta a_i \right|_r
              \left(\frac{e_{\res}}{\mu_j} \right)^2,
\label{eq:tau_tp}
\end{equation} 
where the mass and orbital state of the perturbing
planets are encapsulated in $\mu_j$ (see Equation (\ref{eq:nuj})).  If the
test particle's free eccentricity prior to resonance passage was zero,
then Equation (\ref{eq:eres}) is the new eccentricity after passage of the
secular resonance.  However, when the free eccentricity is not
initially zero, the initial secular coordinates $h_o,k_o$ add
vectorally with those obtained from resonance passage
$h_{\res},k_{\res}$.  The two contributions may interfere constructively
or destructively depending on their relative phase $\gamma$
\citep{Ward_etal_1976,Minton_&_Malhotra_2008} and the resulting
eccentricity can be determined from geometric considerations
\begin{equation}
  e_f = (e_o^2 + e_{\res}^2 + 2e_o e_{\res} \cos \gamma)^{1/2}
  \label{eq:ef}
\end{equation} 
\citep{Heppenheimer_1980}.  Recognizing the phase
between contributions as effectively random and isotropic allows
Equation (\ref{eq:ef}) to be used to as a distribution function of resulting
eccentricities.  For example, given amplitudes for the initial and
resonant eccentricity vectors, $\gamma=0,\pi/2,$ and $\pi$ yield the
maximum, median, and minimum eccentricities expected.

\subsection{Subsystem Approximation for Terrestrial--Giant Planet Interactions}

Due to their much smaller mass and modal amplitudes, the
terrestrial planets have a negligible influence on the evolution of
the giant planets.  The subsystem approximation of \citet{Ward_1981}
treats the terrestrial planets and giant planets as separate and
independently evolving secular systems and then examines their
interaction.  A brief summary of this model is described below
\citep[see][for the complete development]{Ward_1981}.

In this treatment the secular evolution matrix $\mathbf{A}$, whose elements
are defined in Equations (\ref{eq:akk}) and (\ref{eq:aik}), is divided
into submatricies that describe coupling within each terrestrial and giant 
planet subgroup and between the subgroups
\begin{equation}
  \mathbf{A} = \left( \begin{array}{cc}
                    \mathbf{A}_T & \mathbf{P} \\
                    \tilde{\mathbf{P}}   & \mathbf{A}_G
                    \end{array}
               \right).
\end{equation}
Here $\mathbf{A}_{T}$ is the terrestrial $4\times4$ submatrix and
$\mathbf{A}_{G}$ is the $2\times2$ submatrix describing Jupiter and
Saturn.  The direct influence of the giant planets on the terrestrial planets'
precession frequencies is included through the diagonal elements of
$\mathbf{A}_{T}$ (see Equation (\ref{eq:akk})). The coupling between the
terrestrial and giant planet modes are included through  the $4\times2$ and
$2\times4$ off-diagonal submatrices of $\mathbf{A}$ and are labeled
as $\tilde{\mathbf{P}}$ and $\mathbf{P}$. 

The solutions for each subgroup are assumed to be independent.  As
before we label the solution to the $4\times4$ terrestrial subgroup
($\mathbf{A}_{T}$)
\begin{equation}
  \left\{ \begin{array}{c}H\\K\\\end{array} \right\}
  = C_{j} \mathbf{v}_{j} 
  \left\{ \begin{array}{c}\sin (g_jt+\beta_j)\\ \cos (g_jt+\beta_j)\\\end{array} \right\}
\end{equation}
with $j=1\mbox{--}4$.  Similarly, the solution for the Jupiter-Saturn subgroup
($\mathbf{A}_{G}$) is
\begin{equation}
  \left\{ \begin{array}{c}H\\K\\\end{array} \right\}
  = D_n \mathbf{u}_{n} 
  \left\{ \begin{array}{c}\sin (\nu_nt+\delta_n)\\ \cos (\nu_nt+\delta_n)\\\end{array} \right\},
\end{equation}
where $\nu_n$ and $\mathbf{u}_n$ are the two-planet eigenfrequencies
and eigenvectors, $D_n$ and $\delta_n$ are the amplitude and phase of the
two-planet modes, and we retain the index labels $n=5,6$
conventionally used for these modes in the solar system.  Because the
terrestrial planets only weakly affect the precession of the giant
planets, the eigenmodes of the two- and four-planet subgroups are
nearly identical to those of the six-planet system (i.e.,~$\nu_n=g_n$,
for $n=5,6$) shown in Figures \ref{fig:e_vects} and \ref{fig:e_amps}.
Hereafter we refer to the giant planet eigenfrequencies with the
symbol $\nu_n$.  

\begin{figure}[!ht]
\begin{center}
\resizebox{0.40\textwidth}{!}{\includegraphics{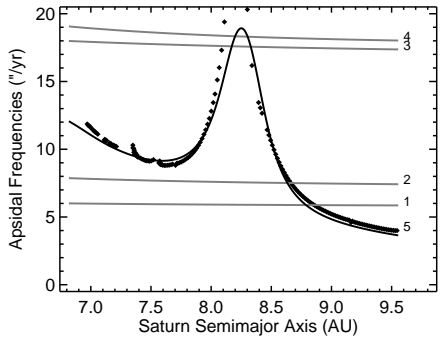}}
\end{center}
\caption{Eigenfrequencies computed using the subgroup approximation
  as a function of Saturn's semimajor axis.  Each frequency is
  labeled with its mode index $j$ at the right.  Jupiter is
  located at 5.2 AU.
  The $j=1\mbox{--}4$ modes that describe the terrestrial planets are
  shown with gray lines.  The $\nu_5$ frequency computed with modified
  secular theory secular is shown with a black solid line. Black
  diamonds show the FMFT measurements of the $\nu_5$ from a series of
  $N$-body simulations.  The $\nu_5$ mode, associated 
  with Jupiter, resonates with each of the terrestrial modes during
  the divergent migration of Jupiter and Saturn.  
  \label{fig:gjvsas}}   
\end{figure}

\subsection{The Sweeping $\nu_5$ Resonance during Migration}

Figure \ref{fig:gjvsas} shows the four eigenfrequencies of the
terrestrial subgroup along with the $\nu_5$ frequency as a function of
Saturn's semimajor axis.  As before, Jupiter is held fixed near its
current orbit at 5.2 AU.   The terrestrial frequencies are
predominantly determined by their  mutual interactions and Jupiter's
proximity and are only weakly influenced by Saturn's orbital position.
This is illustrated in Figure \ref{fig:gjvsas} as the terrestrial
frequencies ($g_{1\mbox{--}4}$ shown in gray) are nearly constant as Saturn's
orbital radius expands.

The $\nu_5$ frequency decreases from about 12\arcsec\mbox{yr}$^{-1}$ to about
8.5\arcsec\mbox{yr}$^{-1}$ as Saturn's orbit expands from just outside
the 3:2 mean motion resonance with Jupiter to about 7.8 AU.  The small
gaps in the FMFT measurement for the $\nu_5$ frequency when Saturn is
at 7.3 AU and 7.55 AU result from the dispersion of forcing
frequencies introduced by the nearby 5:3 and 7:4 MMRs between Jupiter
and Saturn.  As Saturn's orbit expands beyond 7.8 AU toward the 2:1
with Jupiter near 8.25 AU,
the influence of the 2:1 MMR dominates over the linear secular terms
and the $\nu_5$ frequency increases from
8.5$\arcsec\mbox{yr}^{-1}\rightarrow20\arcsec\mbox{yr}^{-1}$. It resonates
with the $g_3$ and $g_4$ terrestrial frequencies when Saturn is near
8.17 AU and 8.20 AU, respectively.  

As Saturn's orbital radius is increased outside the 2:1, the effect of
the mean motion resonance weakens with distance and the $\nu_5$
frequency again decreases from about 20$\arcsec\mbox{yr}^{-1}$ to
4$\arcsec\mbox{yr}^{-1}$ as Saturn's orbit expands from 8.3 AU to 9.54
AU.  The $\nu_5$ resonates a second time with both the $g_4$ and $g_3$
just exterior to the 2:1 MMR when Saturn is near 8.30 AU and 8.33 AU.
As Saturn's orbit is 
expanded further, the $\nu_5$ resonates with the $g_2$ and $g_1$ when
Saturn is near 8.63 AU and 8.79 AU.  In total we have identified six
individual secular resonances between the $\nu_5$ frequency and a
terrestrial eigenmode that occurs as a result of the divergent
migration, of Jupiter and Saturn, from just outside their mutual 3:2
mean motion resonance to their present orbits.   The semimajor axes of
Saturn for each of these resonances is listed in Table
\ref{tab:secres}.  In their numerical exploration of this problem,
\citet{Brasser_etal_2009} parameterized the giant planet  configuration
at the time of resonance with the ratio of the orbital periods of
Saturn to Jupiter ($P_S/P_J$).  Because the value of the period ratio
($P_S/P_J$) at $g_j-\nu_5$ resonances is weakly dependent on Jupiter's
semimajor axis this ratio can be used as a convenient independent
variable to approximately identify resonant configurations.  Its value
for each secular resonance is listed in Table \ref{tab:secres}.

\subsection{Secular Resonances, ``Speed Traps'' and Critical
  Migration Parameters}

Using the subgroup approximation, \citet{Ward_1981} showed that
tuning the system through a linear secular resonance between the
terrestrial and giant planets (i.e.,~$\nu_n=g_j$) excites the terrestrial mode
amplitude by an amount
\begin{equation}
C_{j,\res} = D_n F_{jn} \left( \frac{2\pi}{|\Delta \dot{g}_{jn}|}
\right)^{1/2},
\label{eq:cjres}
\end{equation}
where $\Delta \dot{g}_{jn} = \dot{g}_j - \dot{\nu}_n$.  Here $F_{jn}$
is an effective width of the resonance given by
\begin{equation}
  F_{jn}=\tilde{\mathbf{v}}_j \cdot \mathbf{P} \cdot \mathbf{u}_n,
\end{equation} 
where $\tilde{\mathbf{v}}_j$ is the transpose of $\mathbf{v}_j$.
Equation (\ref{eq:cjres}) is the generalization of the test particle 
response of resonance passage (Equation (\ref{eq:eres})) to a four-planet
terrestrial subsystem.  Here the mode amplitudes $C_j,D_n$
take the place of modal eccentricities, and the functional dependence
on the migration rate through $\Delta \dot{g}_{jn}$ is identical to
the test particle case. 
 
The AMD exchanged with the terrestrial planets through the resonance is
\begin{equation}
   C_{j,\res}^2 = D_n^2 \tau \left[2\pi F_{jn} \left|\sum_i \left(\frac{dg_j}{da_i} -
                  \frac{d\nu_n}{da_i} \right) \Delta a_i
              \right|^{-1}_r \right].
\label{eq:amdres}
\end{equation}
The excitation of the terrestrial mode depends on three principal
factors: the configuration of the planetary system at resonance
(determined by the value of the terms in square brackets), the
amplitude of the resonant giant planet mode ($D_5$), and the migration
timescale ($\tau$).   

It is convenient to combine the forcing amplitude and migration
timescale into a single migration parameter 
\begin{equation}
  \hat{\tau} = \left(\frac{D_5}{D_{5,\obs}}\right)^2 \tau = \left( \frac{e_{55}}{0.0433} \right)^2 \tau.
\label{eq:mp}
\end{equation}
The value of the migration parameter is proportional to the AMD
exchange and can be used to identify the family of giant planet
eccentricity states (e.g.,~$e_{55}$) and migration timescales ($\tau$)
that produce a certain level of excitation.  In this
definition, we have scaled the migration parameter so that
$\hat{\tau}=\tau$ when Jupiter and Saturn migrate with eccentricities
equal to their present-day values. 

Using the above relations we identify the critical migration
parameter for each resonance ($\hat{\tau}_{jn}^*$) capable of exciting a 
terrestrial mode to its observed amplitude 
\begin{equation}
  \hat{\tau}_{jn}^* = \frac{1}{2 \pi F_{jn}^2} \left( \frac{C_{j,\obs}}{D_{n,\obs}}\right)^2 
              \left|\sum_i \left(\frac{dg_j}{da_i} -
                  \frac{d\nu_n}{da_i} \right) \Delta a_i \right|_r .
\label{eq:tau_mode}
\end{equation} 
In a sense, passage through the secular resonance may
be used as a dynamical ``speed trap'' to constrain the eccentricity
state and maximum migration timescale (or minimum migration rates) of
Jupiter and Saturn at the time of resonance.      For each secular
resonance we have computed the giant planet orbits, resonance widths
($F_{jn}$), and critical migrations parameters ($\hat{\tau}_{jn}^*$)
for exciting the resonant terrestrial   mode to its observed
amplitude.  The characteristics describing these resonances  are
listed in Table \ref{tab:secres}.  Equation (\ref{eq:cjres}) indicates
that resonant excitation is  dependent on the slope $d\nu_5/da_6$.
The $g_{3}-\nu_5$ and  $g_{4}-\nu_5$ secular resonances occur when
Jupiter and Saturn are very near the 2:1 MMR and the modified secular
model becomes less accurate.  For these, we use values for the $\nu_5$
frequency gradient interpolated from the FMFT results in
Figure \ref{fig:gjvsas}. 

The effective width of a resonance ($F_{jn}$) is determined by the
projection of the resonant terrestrial eigenvector ($\mathbf{v}_j$)
onto the off-diagonal matrix $\mathbf{P}$ that accounts for coupling
between the terrestrial and giant planet subgroups.  As these terms
scale with planet mass (see, e.g., Equation (\ref{eq:aik})), the effective
width tends to be larger for resonances with modes where more massive
planets have larger eigenvector components ($v_{kj}$).  The resonance
widths are shown in Table \ref{tab:secres}.   Since the $j=2,\,3$ modes
have the largest eigenvector components for Earth and Venus (see
Figure \ref{fig:e_vects}) resonances with these modes have greater
effective widths than resonances with the $j=1,\,4$ modes.    

Also, the greater effective widths of the resonances with the $j=2,\,3$
modes and  Equation~(\ref{eq:cjres}) indicate that for a given frequency
sweeping rate at resonance ($\Delta \dot{g}_{jn}$),  the $j=2,\,3$ modes
experience a greater excitation of their amplitudes than the $j=1,\,4$
modes.  This susceptibility to excitation appears in contrast to the
mode amplitude partitioning of the terrestrial planets, where $C_2$
and $C_3$ are smaller than $C_1$ and $C_4$ (see
Figure~\ref{fig:e_amps}). 

This general model can be used to predict the excitation of modal
eccentricity amplitudes due to passage through a secular resonance ($
e_{ij,\res} $)
\begin{equation}
  e_{ij,\res} =
  e_{ij,\obs}\left(\frac{\hat{\tau}}{\hat{\tau}_{j5}^*}\right)^{1/2},
\label{eq:erscale}
\end{equation}
where $e_{ij,\obs}$ are the eccentricity amplitudes of the present-day
system listed in Table \ref{tab:eij}.
Note that in both the test particle and subgroup approximation the
resonant forcing of eccentricity scales as $\hat{\tau}^{1/2}$ (or
equivalently $e_{\res} \propto  e_{55} \tau^{1/2}$).

These relations and the observed state of the terrestrial planets  can
be used to constrain the giant planet migration parameter at the time
of passage through a secular resonance.   
For example, if we know Jupiter's eccentricity at the time of
resonance (e.g.,~the current value) then we can constrain the migration
timescale of Jupiter and Saturn (and vice versa).   

\begin{table}[!h]
\begin{center}
\footnotesize

\begin{tabular}{cccccccc}

\hline 
\multicolumn{2}{c}{Mode} & Saturn & Period Ratio & Frequency & Resonance Width &
Critical Migration Parameter\\ 

$j$ & $n$ &  $a_S$(AU) & $P_S/P_J$& $g_j$($\arcsec$yr$^{-1}$) &
$F_{jn}$($\arcsec$yr$^{-1}$) & $\hat{\tau}^*_{jn}$ (Myr) \\ \hline
 3  & 5 &  8.17  & 1.968 & 17.59  & \hspace{1ex}0.040 & 0.598 \\
 4  & 5 &  8.20  & 1.979 & 18.34  & --0.028  &           3.609 \\
 4  & 5 &  8.30  & 2.016 & 18.30  & --0.025  &           1.840 \\
 3  & 5 &  8.33  & 2.025 & 17.57  & \hspace{1ex}0.036 & 0.329 \\
 2  & 5 &  8.63  & 2.138 &  7.51  & --0.052  &           0.050 \\
 1  & 5 &  8.79  & 2.196 &  5.87  & \hspace{1ex}0.013 & 0.683 \\
\hline
\normalsize
\end{tabular}
\end{center}
\caption{
Secular Resonances During the Divergent Migration
  of Jupiter and Saturn (See
  e.g.,~Figure~\ref{fig:gjvsas})\tablenotemark{\, }  
\label{tab:secres} } 
\tablenotetext{\, }{\textbf{Note.} In this  model Jupiter's semimajor
  axis is constant at 5.2 AU.  }
\end{table}


The critical migration parameters $(\hat{\tau}^*_{jn})$ required to
excite the terrestrial modes to their observed amplitudes are listed
in Table \ref{tab:secres}.   All of these migration parameters are
shorter than the migration timescales observed in self-consistent
models of planetesimal-driven giant planet migration
\citep[e.g.,~$\tau\sim$5--20Myr;][]{Hahn_&_Malhotra_1999,Gomes_etal_2004}
by a factor ranging from a few to 200.  These secular resonances
constrain migration to occur with a smaller $e_{55}$ Jovian
eccentricity amplitude and/or short  migration timescales to avoid
driving the terrestrial eccentricities to large values.     Further,
we note that resonances between the $\nu_5$ frequency and the $g_2$
and $g_3$ terrestrial modes, predominantly responsible for the small
eccentricities of Earth and Venus, place the strongest constraints on
the migration of Jupiter and Saturn.    We compare these model
predictions with the results of orbital integrations below.

\section{Jupiter and Saturn's Migration Near the 2:1 MMR}
\label{sec:JS12}

As Jupiter and Saturn diverge across their mutual 2:1 MMR they perturb
the terrestrial planets via at least five distinct dynamical events.
The $g_{3}-\nu_5$ and the $g_{4}-\nu_5$ resonances are each
encountered on both sides of the 2:1 MMR.   In addition, the 2:1 MMR
excites the eccentricity of Jupiter and Saturn on a timescale
comparable to the circulation period of the resonant argument.  Near
the 2:1 MMR this is $\mathcal{O}(10^3)$ yr.  Because the typical
secular   timescales of the terrestrial planets are more than an order
of magnitude longer ($\gtrsim\mathcal{O}(10^4)$), exciting the
eccentricities of Jupiter and Saturn in this way can be thought of as
rapid, non-adiabatic changes in the $C_{5}$ and $C_{6}$ secular modal
amplitudes.  

To clarify the implications of individual events our analysis
separates the $g_{3}-\nu_5$ and $g_{4}-\nu_5$ secular resonances, that
occur as Jupiter and Saturn migrate near the 2:1, from the
$g_{2,1}-\nu_5$ resonances, that occur for larger Jupiter--Saturn
separations.  We analyze the dynamics of these individual resonances
using both $N$-body simulations and secular theory and examine how they
act in concert.

\subsection{Initial Conditions}

To clearly illustrate the role of these events using orbital
integrations it is useful to start with terrestrial mode amplitudes
($C_{1\mbox{--}4}$) that are initially very small.  The initial conditions for
these simulations were prepared in the following way.  Jupiter was
placed on its current orbit and Saturn at 7.2 AU.  Both giant planets
were given initial eccentricities of 0.01 and comparable inclinations.
The terrestrial planets were given their current semimajor   axes,
small eccentricities, and inclinations ($e=0.01$) and randomly chosen
angular phases.  This configuration was processed further by
integrating it forward in time while applying an additional force to
damp the eccentricities and inclinations of the terrestrial planets
with a decay timescale of three million years.  The timescale was
chosen to be slow compared to the secular timescale so that it
effectively damped the terrestrial mode amplitudes rather than the
elements of a single planet \citep{Agnor_&_Ward_2002}.  For these
initial conditions the FMFT-measured eccentricity amplitudes are
$e_{ij}\lesssim0.005$ for $i,j=1\mbox{--}4$ (for the terrestrial planets) and
$e_{55}\simeq0.015$ for Jupiter, about one-third its observed
value. This value of $e_{55}$ is comparable to that suggested in
\citet{Tsiganis_etal_2005}.   For these initial conditions, the
critical migration timescales needed to excite the terrestrial modes
to their observed values are about a factor of nine longer than the
migration parameter listed in Table \ref{tab:secres} (see
Equation~(\ref{eq:mp})). 

We have performed a series of $N$-body orbital integrations using this
initial state.   In these models Saturn is forced to migrate outward
across the 2:1, halting its migration at 8.5AU.  The migration force
applied produces exponential migration with an $e$-folding timescale
($\tau$) similar to that of \citet{Malhotra_1993}.  The starting and
ending points of Saturn's migration leave the $\nu_5$ frequency
between the $g_2$ and $g_3$ frequencies and avoids depositing the
system near a secular resonance (see, e.g.,~Figure \ref{fig:gjvsas}).
This allows us to isolate the effects of individual events during
migration and use FMFT analysis of the final states to measure the
ultimate eccentricity amplitudes that result. 

In this set of simulations we varied the migration timescale from
$\tau= 10^4\rightarrow 4\times10^7$ years.  The low end of this range
is near the secular timescale and illustrates the response of the
system to rapid, essentially non-adiabatic, evolution.  The upper end
encompasses migration timescales suggested by $N$-body simulations of
planetesimal-driven giant planet  migration
\citep[e.g.,][]{Hahn_&_Malhotra_1999,Gomes_etal_2004}.    

\begin{figure}[!h]
\begin{center}
\subfigure{
\resizebox{0.4\textwidth}{!}{\includegraphics{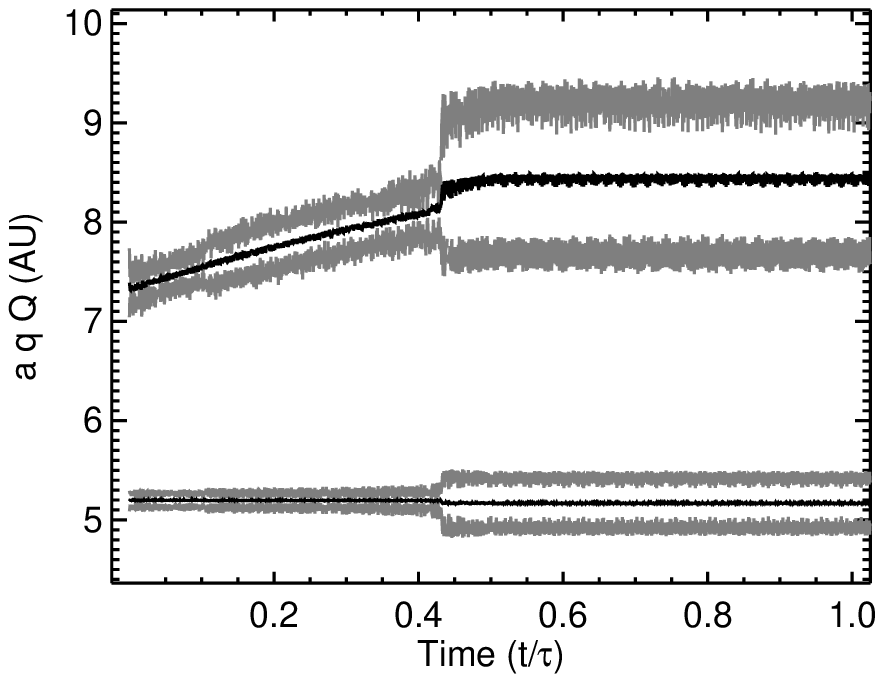}}}\\
\subfigure{
\resizebox{0.4\textwidth}{!}{\includegraphics{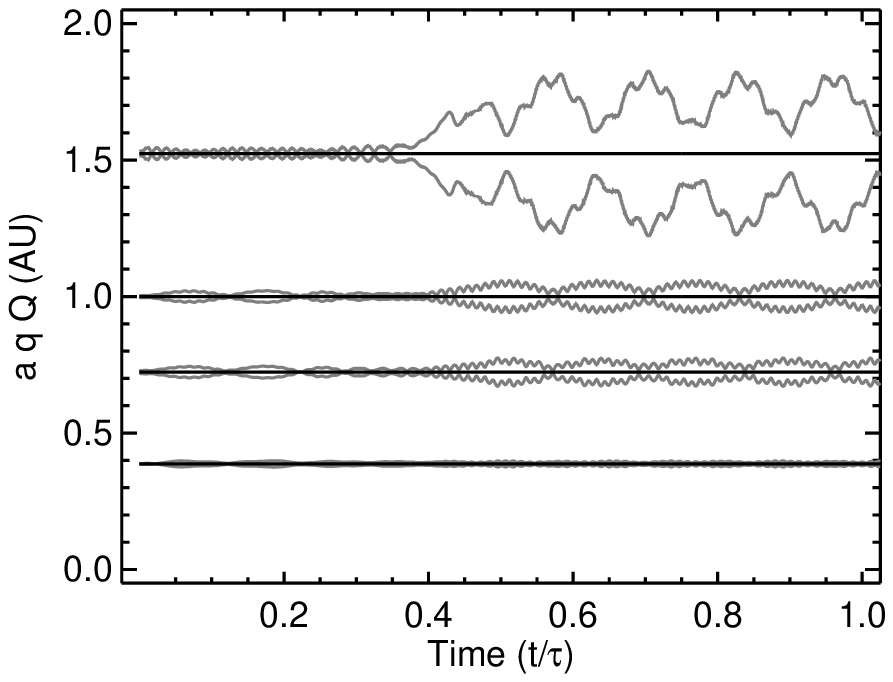}}}

\end{center}
\caption{Forced migration simulation with $\tau = 10$Myr.  The
  semimajor axes of the Jupiter and Saturn and the terrestrial
  planets are shown with a black solid line.  The pericenter and
  apocenter are shown in gray.  The
  excitation of the terrestrial $j=3,\,4$ modes is evident in the
  increased eccentricity of Earth, Venus, and Mars.
 \label{fig:js12_t1e7}}  
\end{figure}

Figure \ref{fig:js12_t1e7} shows an example of the coupling between
the gas giant and terrestrial planets as Jupiter and Saturn are forced
to migrate across their mutual 2:1 mean motion resonance with an
$e$-folding timescale of $\tau=10$Myr.  The top frame shows the
evolution of the semimajor axes, pericenters, and  apocenters of
Jupiter and Saturn.  Their divergent crossing of the 2:1 MMR occurs
near  $t/\tau=0.425$ and the excitation of the giant planet
eccentricities is readily apparent as the increased separation of the
pericenter and apocenter curves from the semimajor axes.  The lower
panel shows these same orbital parameters for the terrestrial planets.
Mars' eccentricity is excited to a value similar to the observed one.
As Earth and Venus have significant components in both the $j=3$ and
$j=4$ modes (see, e.g.,~Figure~\ref{fig:e_vects}), their excitation
increases eccentricities of these planets as well.  This excitation is
coincident with Jupiter and Saturn migrating across the 2:1 MMR. 

\subsection{The First Crossing of the $g_3-\nu_5$ and $g_4-\nu_5$ Resonances}
\begin{figure}[!h]
\begin{center}

\subfigure[Mars]{
\resizebox{0.40\textwidth}{!}{\includegraphics{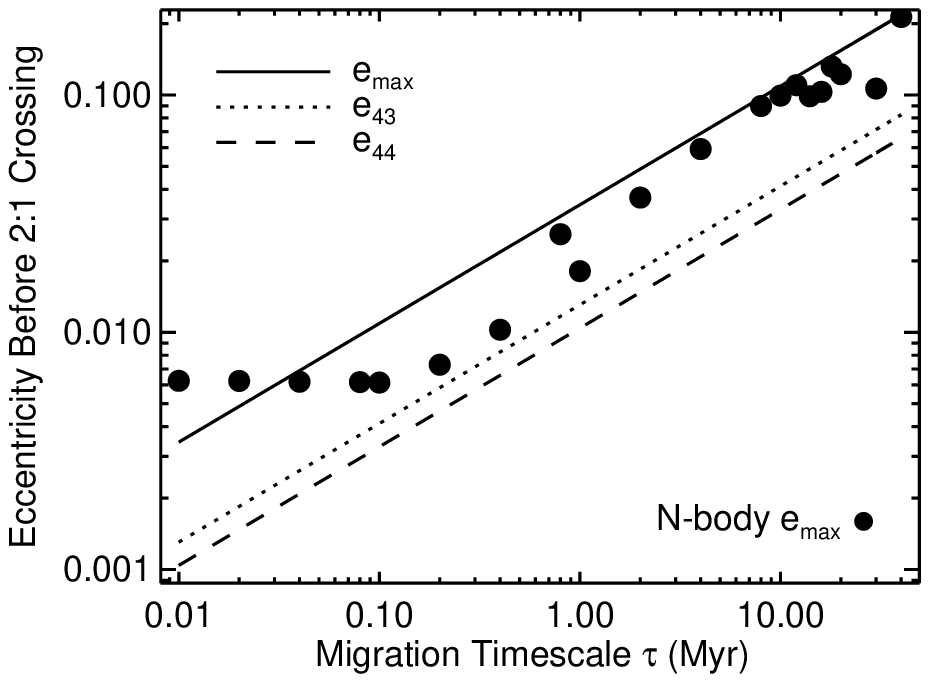}}}
\subfigure[Earth]{
\resizebox{0.40\textwidth}{!}{\includegraphics{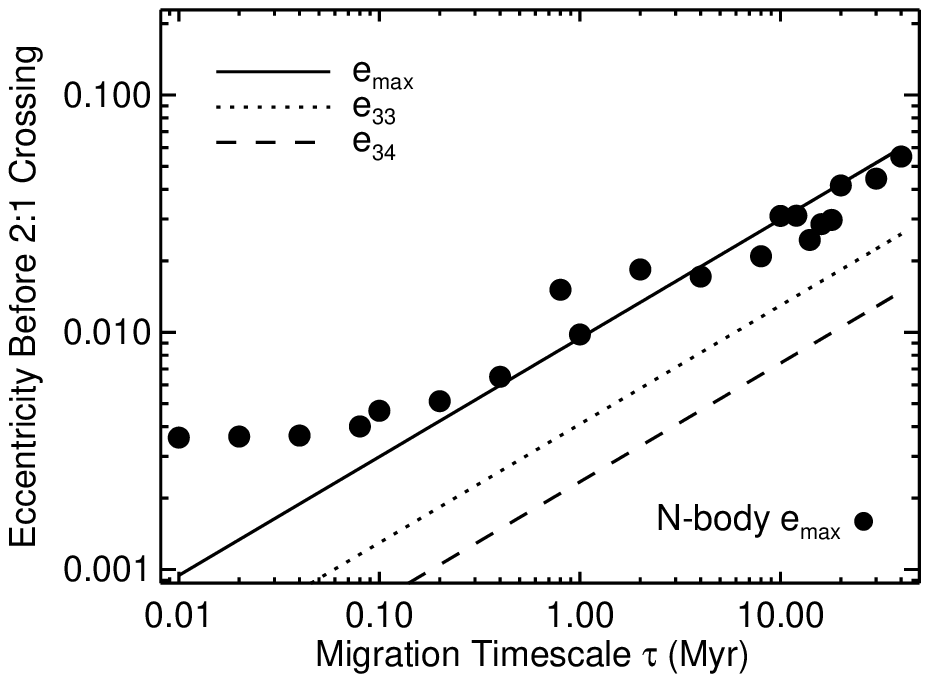}}}

\end{center}
\caption{Eccentricity of Mars ($i=4$) and Earth
  ($i=3$) excited during Jupiter and Saturn's approach to their 2:1 mean
  motion resonance as a function of the migration timescale ($\tau$).
  The dotted and dashed lines indicate the  
  prediction of resonant excitation of the $e_{i3}$ and
  $e_{i4}$ eccentricity amplitudes.  The theoretical maximum
  eccentricity that might result is the sum of these two contributions
  and is shown with the solid line.  The maximum osculating
  eccentricity observed in the simulations prior to 2:1 MMR crossing
  is shown with a filled circle. In both the linear
  theory and the $N$-body simulations the forcing eccentricity of Jupiter
  is about one-third the observed value (i.e.,~$e_{55}\approx0.015$).
  The critical migration parameters $\hat{\tau}_{35}^*$ and  $\hat{\tau}_{45}^*$ suggest migration
  timescales of $\tau =5
  \mbox{ and } \tau =30$ Myr can excite the $j=3$ and $j=4$ modes to
  their observed amplitudes.
  \label{fig:emax_pre21}}  
\end{figure}

Figure \ref{fig:emax_pre21} shows the eccentricity excitation due to
the first passage through the
$g_3-\nu_5$ and  $g_4-\nu_5$ secular resonances as a function of the
migration timescale $\tau$.  The theoretical predictions for the $j=3$
and $j=4$ eccentricity amplitudes of Earth and Mars computed with Equation (\ref{eq:erscale}) are shown with dotted and dashed lines.
The maximum eccentricity for planet $i$ results when the contributions
from these two modes are in phase, yielding a value that is simply their sum
$e_{i,\max}=e_{i3}+e_{i4}$.  This is shown with a solid line for
both planets.  We compare these predictions with the results of
$N$-body integrations by identifying the maximum osculating eccentricity of
Mars and Earth observed in each simulation prior to the time when Jupiter
and Saturn crossed their 2:1 MMR.  These values are shown with filled
circles.    

For short migration timescales ($\tau\lesssim 10^5$ yr), the initial
eccentricities ($e_{ij}\simeq 0.005$) dominate over contributions from
the secular resonances.  For longer migration timescales, the excitation of the
resonances is more apparent.  The $N$-body results always fall below the
theoretical maximum and above the predicted amplitudes due to single
modes.  The time required to reach the maximum eccentricity of a
two-mode system  is the about half the beat period of two
eigenfrequencies.  For the $j=3$ and $j=4$ modes this is $\sim$1 Myr.
For $\tau\gtrsim 1$Myr there is sufficient time in the simulation to
realize the maximum eccentricity prior to the Jupiter--Saturn 2:1
crossing and the values measured from $N$-body simulations cluster
tightly near the theoretical maximum.  For $\tau\gtrsim 10^{5}$ yr the
maximum  eccentricities of both planets follow the $\tau^{1/2}$
scaling of Equation~(\ref{eq:erscale}) and are within 0.005 (the initial
amplitude) of the theoretical prediction.  Clearly, the theory
described above provides an accurate description of excitation during
first passage through the $g_3-\nu_5$ and $g_4-\nu_5$ secular resonances.

We note that the predicted excitation of both modal eccentricity
amplitudes of Mars ($e_{43}$ and $e_{44}$) are comparable (see Figure
\ref{fig:emax_pre21}(a)). For a given migration timescale $\tau$, the rate of 
frequency sweeping through the $g_3-\nu_5$ resonance is less (see
$|d\nu_5/da_6|$ in Figure~\ref{fig:gjvsas}), and the effective resonant width
($F_{35}$) greater, than for sweeping through the $g_4-\nu_5$
resonance.  Both factors contribute to the greater excitation of $C_3$
relative to $C_4$.  Finally, the eigenvector components of Mars in the
$j=3$ and $j=4$ modes are comparable in magnitude
(i.e.,~$|v_{4,3}|\simeq |v_{4,4}|$, see Figure~\ref{fig:e_vects}). 
The net effect is that for any migration timescale ($\tau$), the
eccentricity excitation of the $j=3$ component in Mars'
eccentricity is actually \emph{greater} than that of the $j=4$ mode
(i.e.,~$e_{43} \simeq 1.1 e_{44}$, see Figure~\ref{fig:emax_pre21}(a)).
This appears in contrast to the observed eccentricity partitioning for
Mars, where $e_{43} = 0.020$ and $e_{44}=0.067$.   

\subsection{Crossing the 2:1 MMR}
The increase in the eccentricities of Jupiter and Saturn as
they diverge across their mutual 2:1 resonance might be communicated
to the terrestrial planets.  This change in
the eccentricities and longitude of perihelia occurs over the
circulation period of the 2:1 resonant argument ($\mathcal{O}(10^3)$ yr).
As this change occurs over a 
timescale much shorter than the secular periods, this eccentricity
excitation might be viewed as a sudden change in system's
mode amplitudes $C_j$ and phases $\beta_j$.  Examination of the 2:1 MMR crossing
finds that this impulsive event fairly purely excites the $C_6$ mode
amplitude and that the excitation of the $C_5$ mode, is weak
($e_{55}\sim\mathcal{O}(10^{-3})$) and insufficient to explain its
observed value
\citep{Morbidelli_etal_2009}.\footnote{The conference presentation of
  \citet{Cuk_2007} also highlighted this aspect of the 2:1 crossing.}
In general, this resetting of the giant planets' eccentricities should
alter the entire system's mode amplitudes.  
However, in the observed
solar system, the $j=6$ mode weakly contributes to the eccentricity
evolution of the terrestrial planets (see the small values of $e_{i6}$
in Table \ref{tab:eij}).  So, the terrestrial eccentricity amplitudes
($e_{ij}$, where $i,j =1\mbox{--}-4$) are modified by small amounts to
accommodate an increase in the $j=6$ amplitude.  On this basis, we
suggest that the impulsive excitation of the $C_6$ mode amplitude is
only weakly communicated to the terrestrial planets and its effect on
the terrestrial $j=1\mbox{--}4$ modes is small relative to contributions from
the $g_j-\nu_5$ secular resonances.    

\subsection{The Second Crossing of the $g_3-\nu_5$ and $g_4-\nu_5$ Resonances}
The excitation obtained from the second crossing of the $g_{4}-\nu_5$
and $g_{3}-\nu_5$ resonances can also be predicted using Equation~(\ref{eq:erscale}).    For a given migration timescale the
excitation of the $j=3,\,4$ resonances encountered inside of the
Jupiter--Saturn 2:1 MMR are comparable to the excitation 
due to these secular resonances outside the 2:1.

\begin{figure}[h!]
\begin{center}

\subfigure[$\tau=20$Myr - Constructive Interference]{
\resizebox{0.40\textwidth}{!}{\includegraphics{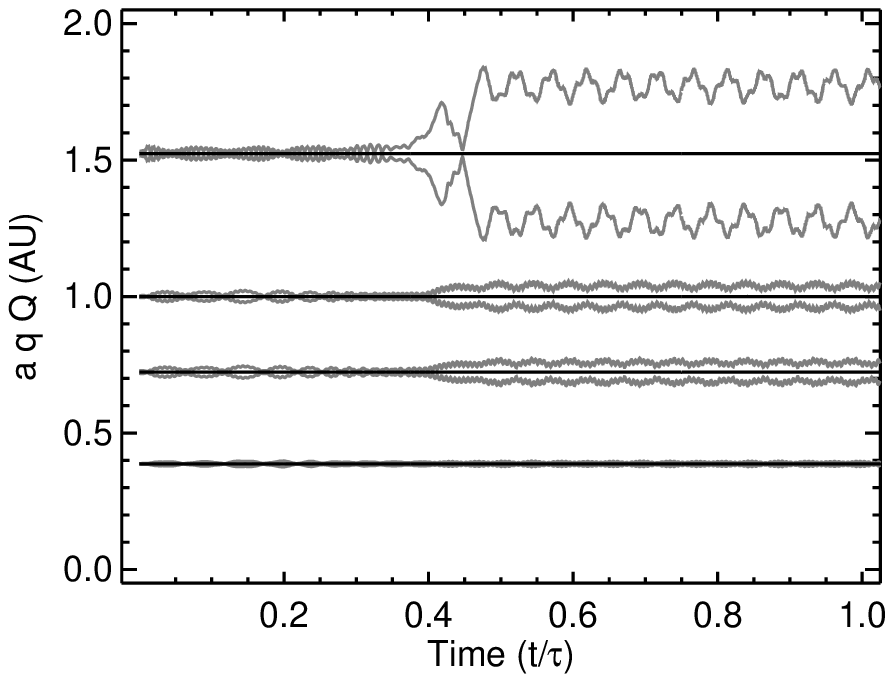}}}
\subfigure[$\tau=40$Myr - Destructive Interference]{
\resizebox{0.40\textwidth}{!}{\includegraphics{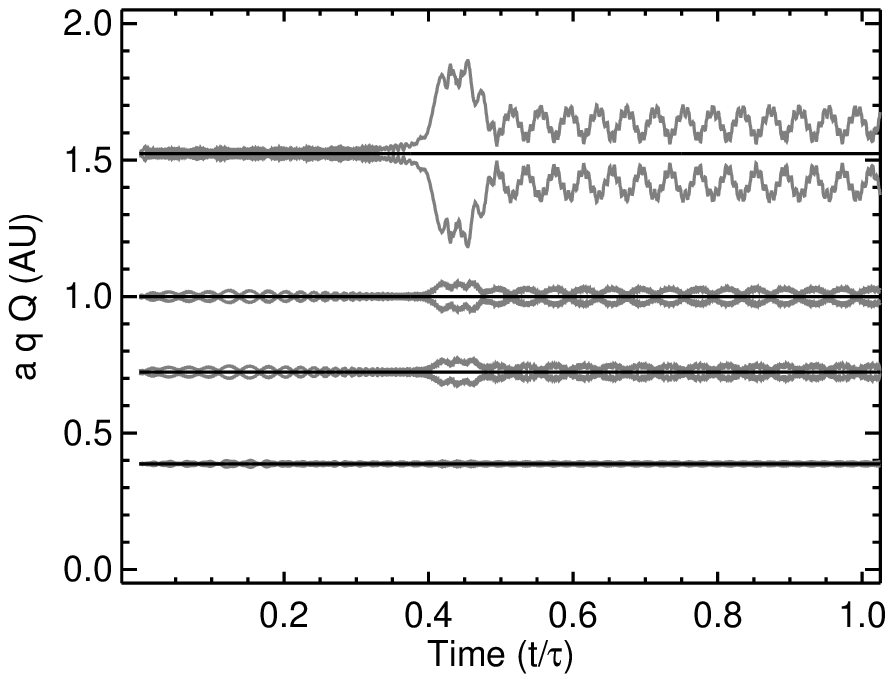}}}


\end{center}
\caption{Forced migration simulations with $\tau = 20 \mbox{ and }
  40$ Myr.  The evolution of the Jupiter and Saturn (not shown here)
  is similar to that in Figure \ref{fig:js12_t1e7}.  The black and
  gray lines represent the semimajor
  axes, pericenter, and apocenter distances, respectively.  In (a) the
  contributions from crossing the secular resonances twice has an additive effect, resulting in a
  large final eccentricity for Mars. In (b) the excitation of
  Mars, Earth, and Venus due to two passages through the
  $g_{3}-\nu_5$ and $g_{4}-\nu_5$ resonances partially
  cancel and leave Mars, Earth, and Venus in a dynamical state similar
  to the observed one.
  Because of the smaller Jovian eccentricity amplitude
  $e_{55}\simeq0.015$ the simulation in (a) and (b) correspond to migration
  parameters of 2.4 and 4.8 Myr, respectively.  
  \label{fig:interfere}}    
\end{figure}

Depending on the relative phasing between excitation from the first and
second resonance crossings, these two contributions may interfere
constructively or destructively and result in eccentricities larger or
smaller than predicted for passage through a single resonance.  Figure
\ref{fig:interfere} shows examples from $N$-body simulations that
illustrate both constructive and destructive interference due to successive
passages through the $g_{3} - \nu_5$ and  $g_{4} - \nu_5$ resonances
as Jupiter and Saturn cross their 2:1 MMR.   

In both examples the migration timescale $\tau$ is sufficiently long
that the first secular resonance crossing near $t/\tau=0.4$ excites
both $j=3$ and $j=4$ modes to amplitudes larger than their observed
values.  For the case shown in (a) the second passage through
resonance results in a further increase in eccentricity amplitudes.
In the second example (b), the contributions from the two resonances interfere
destructively resulting in a final eccentricity state consistent with the
observed terrestrial planets.  This occurs despite a migration
parameter greater than the critical value. 

The latter result illustrates that Jupiter and Saturn's crossing 
of the 2:1 MMR can temporarily excite the eccentricity of Mars to values
of 0.2 or larger.  If Mars' orbit was excited in this way
it may have facilitated the depletion of otherwise dynamically
stable and unpopulated regions exterior to Mars' $a=1.7\mbox{--}2.0$ AU
\citep{Evans_&_Tabachnik_2002,Bottke_etal_2010}.

\begin{figure}[!h]
\begin{center}
\subfigure[Earth]{
\resizebox{0.40\textwidth}{!}{\includegraphics{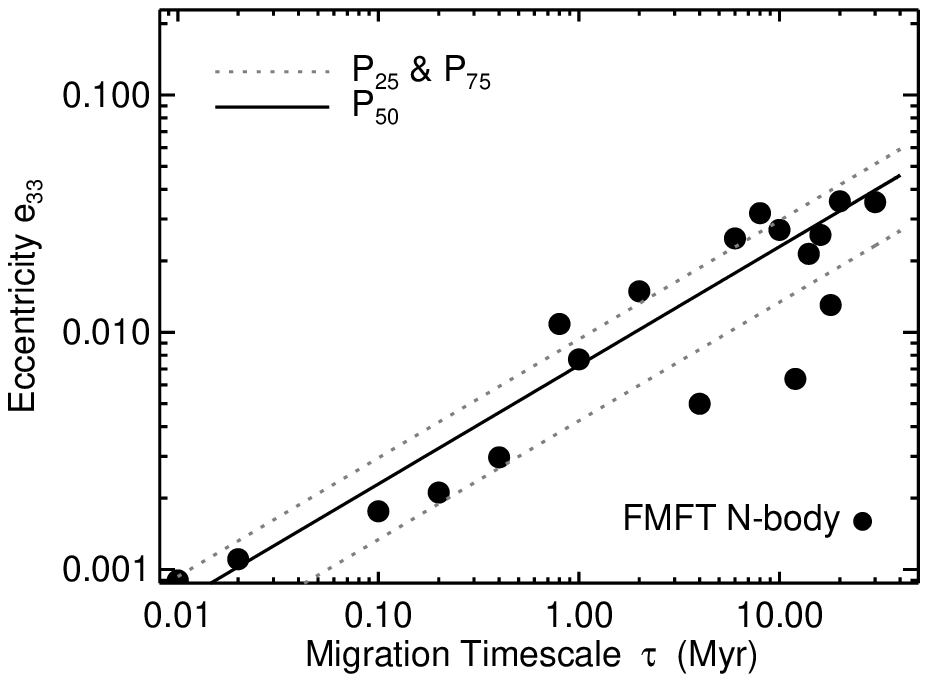}}}
\subfigure[Mars]{
\resizebox{0.40\textwidth}{!}{\includegraphics{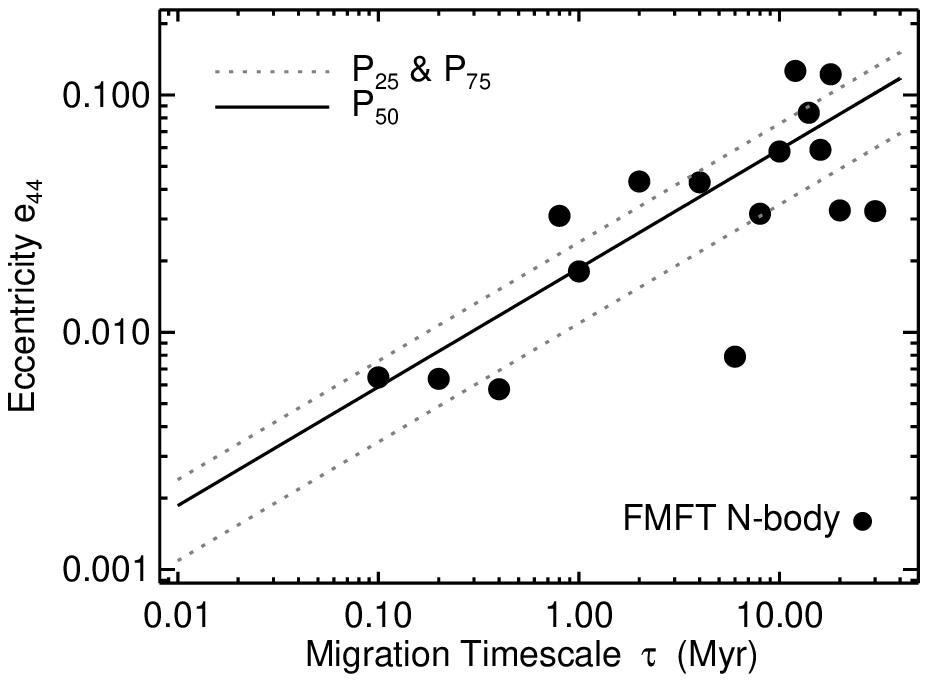}}}

\end{center}
\caption{FMFT-measured eccentricities for Earth ($e_{33}$) and
  Mars ($e_{44}$) are shown as a function of the migration timescale
  ($\tau$).  The scatter in these measured results is due to
  interference between contributions from multiple secular resonance
  passages.  Using secular theory and the distribution function of
  Equation~(\ref{eq:ef}) we also show the eccentricities of the 50th (solid
  line), 25th, and 75th percentiles
  ($P_{25}$ and $P_{75}$ dotted lines).  For comparison the principal
  eccentricity amplitudes of Earth and Mars are $e_{33}\simeq 0.01$ and
  $e_{44}\simeq0.07$, respectively (see
  Table~\ref{tab:eij}).  \label{fig:dekj_js21}}   
\end{figure}

For each simulation in this set of models we conducted an additional
simulation to measure the eccentricity 
amplitudes ($e_{ij}$) of the terrestrial planets via Fourier
analysis.  The FMFT-measured values for Earth ($e_{33}$) and Mars ($e_{44}$) are
shown in Figure \ref{fig:dekj_js21} with filled circles as a function of the
migration timescale $\tau$.  Using linear secular theory we have
computed the eccentricity amplitude for each planet due to secular
resonance crossing both interior and exterior to the 2:1 MMR.  As
expected the net effect of the contributions from these two resonances
is stochastic and depends on the relative phase between them.  Using
Equation~(\ref{eq:ef}) as a distribution function we show the median
eccentricity expected with a black solid line.   The eccentricity
amplitudes of the 25th and 75th percentiles ($P_{25}$ and $P_{75}$)
are shown with dotted lines.  Each line follows the $\tau^{1/2}$ scaling.

Due to the stochastic influence of multiple resonance passages there is
significant scatter in the eccentricities produced.  However, about
half of the FMFT-measured eccentricity amplitudes lie between the
25th and 75th percentile curves indicating that the results are in
good accord with the distribution suggested by Equation~(\ref{eq:ef}) and
demonstrate that the linear model effectively predicts characteristic
levels of excitation and interference. 

\subsection{Interference of Multiple Contributions}
\label{sec:interfere}
If the $\nu_5$ frequency swept through resonances with both the $g_3$
and $g_4$ frequencies then the large amplitude of the $j=4$ mode
relative to the $j=3$ mode requires explanation.  Perhaps the relative
amplitudes of the $j=3,4$ modes resulted from the interference between
multiple contributions.  Linear secular theory and our results above
suggest that the final modal eccentricity amplitudes $e_{ij}$ can be
calculated as the vector sum of multiple contributions (as in 
Equation~\ref{eq:ef}).  The $g_3-\nu_5$ and  $g_4-\nu_5$ resonances are close
together and are each encountered twice in rapid succession as Jupiter
and Saturn diverge across their 2:1 MMR.   These resonant
contributions combine with any primordial amplitude that
resulted from the planet formation epoch.   

Studies of the gas free, late stage accretion of terrestrial planets
have had difficulty simultaneously accounting for the small
eccentricities of Earth and Venus and the smallish masses of Mercury
and Mars.   Further, terrestrial planet formation
simulations often produce planets with masses 0.7-1.0$M_{\oplus}$ with
orbital eccentricities larger than Earth or Venus \citep[see,
e.g.,][]{Chambers_&_Wetherill_1998,Agnor_etal_1999,Chambers_2001,Obrien_etal_2006,Raymond_etal_2009}
and suggest that a broad range of eccentricities might plausibly arise
from accretion processes. 

Here we examine the relative probability of eccentricity amplitudes
combining to yield a dynamical state
comparable to the observed one for the terrestrial planets.  We
consider the contributions to each mode individually below.
We compute the final value of the $e_{33}$ eccentricity
amplitude from the vector sum of an assumed initial amplitude (e.g.,~one
resulting from formation) and two contributions from passage through
the $g_3-\nu_5$ resonances on both sides of the 2:1 Jupiter--Saturn
MMR.  For a given value of the migration
parameter ($\hat{\tau}$), the amplitudes of resonant contributions are
determined using Equation~\ref{eq:erscale}.  We then choose their relative
orientations at random, combine them and compute the resultant
eccentricity vector.   For each value of the migration parameter this
procedure is repeated $10^4$ times and the cases where the final
eccentricity amplitude is similar to the observed state of terrestrial
planets are counted.   

When examining the results for the $j=3,\,4$ modes, we consider final
eccentricity states with $e_{33}<0.02$ and  $0.05<e_{44}<0.09$ to be
``close'' to the observed dynamical state of the terrestrial planets.
While this definition is somewhat arbitrary, the first constraint is
within a factor of two of the small $e_{33}$ amplitude, and the second
is within $\simeq30$\% of the observed $e_{44}$ amplitude (see Table 
\ref{tab:eij} for observed values).

\begin{figure}[ht!]
\begin{center}
\subfigure[Percent with $e_{33}<0.02$]{
\resizebox{0.45\textwidth}{!}{\includegraphics{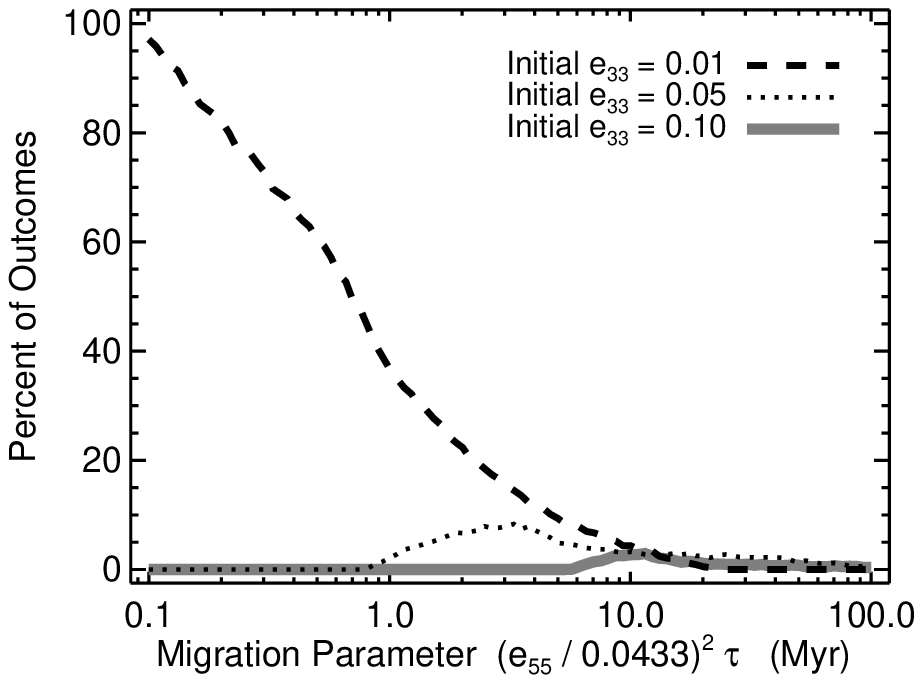}}}
\subfigure[Percent with  $0.05 < e_{44}< 0.09 $]{
\resizebox{0.45\textwidth}{!}{\includegraphics{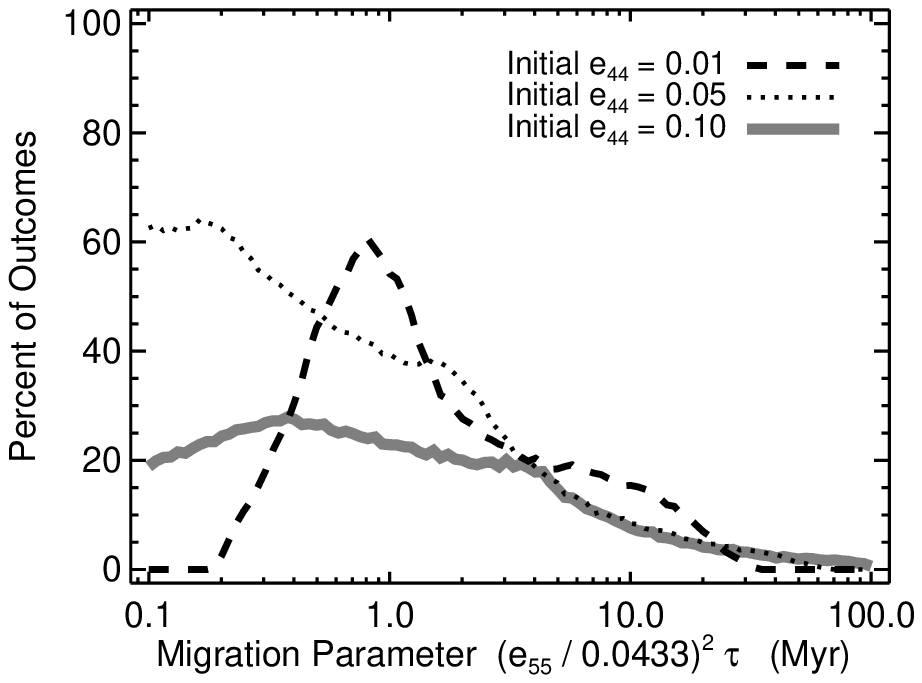}}}
\end{center}

\caption{Percent of outcomes where the $e_{33}$ and $e_{44}$
  eccentricity amplitudes are close to those of the observed terrestrial
  planets are shown as a function of the migration parameter
  (Equation~(\ref{eq:mp})).  As Jupiter and Saturn migrate across the 2:1 MMR,
  the $g_3-\nu_5$ and  $g_4-\nu_5$ resonances are each crossed twice, once
  on each side of the 2:1.   Contributions from these resonances are
  combined with an assumed initial amplitude and the percentages of
  outcomes estimated using a Monte Carlo technique.  Curves for
  different assumed initial eccentricities are shown.
   \label{fig:e33e44}}  
\end{figure}

\subsubsection*{Constraints on the Origin of $e_{33}$}

In Figure \ref{fig:e33e44}(a) we show the percent of cases that
result in $e_{33}<0.02$ as a function of the migration parameter for small
(0.01), modest (0.05), and large (0.10) initial modal eccentricity amplitudes. 
Note that the critical migration parameter for the $g_3-\nu_5$
resonances interior and exterior to the 2:1 are $\hat{\tau}^*_{35}\simeq
0.60 \mbox{ and } 0.33$ Myr, respectively.  For migration parameters
larger than 0.7 Myr, crossing each $g_3-\nu_5$ resonance contributes
a component larger than the observed amplitude and significant
cancellation is required to achieve the observed small amplitude.  For
small initial eccentricity amplitudes (0.01), the percent of outcomes
with $e_{33}<0.02$ decreases from 100\%, to about 20\% as the migration
parameter increases from 0.1 to 2.0 Myr. 

When either the initial amplitude ($e_{33}=0.05, 0.10$) or resonant
contributions are substantially larger than the observed $e_{33}$
value  (e.g.,~for $\hat{\tau}>\hat{\tau}_{35}^*$), significant
cancellation between all three contributions is required to produce a
small final eccentricity.  Figure~\ref{fig:e33e44}(a) shows that strong
cancellation and  $e_{33}<0.02$ can happen, but is an infrequent
occurrence for large initial eccentricity amplitudes
$e_{33}=0.05,\,0.10$ or migration parameters greater than 2--5 Myr.
On the other hand, the origin of the small $e_{33}$ amplitude is
broadly consistent with  a small primordial amplitude
(i.e.,~$e_{33}\simeq0.01$) and migration parameters $\lesssim 1\mbox{--}2$
Myr.\footnote{Note this upper limit on the migration parameter
$\hat{\tau}$ is larger than $\hat{\tau}^*_{35}$ listed in Table
\ref{tab:secres} as we are considering amplitudes larger than observed
amplitudes of $e_{33}$ as acceptably close to the observed system.} It
might also be explained by the  cancellation of large resonant and
initial contributions, but this becomes less likely when the
contributions are substantially larger than the observed amplitude.

\subsubsection*{Constraints on the Origin of $e_{44}$}

As above, we compute the resultant amplitude of the $j=4$ mode as
being comprised of an assumed initial amplitude and contributions from
both $g_4-\nu_5$ resonances.  The critical migration parameters for
the resonances interior and exterior to the 2:1 are
$\hat{\tau}^*_{45}\simeq 3.8 \mbox{ and } 1.7$ Myr, respectively.  The
percent of cases that produce a final eccentricity amplitude with
$0.05<e_{44}<0.09$ is shown in Figure~\ref{fig:e33e44}(b) as a function of
the migration parameter for three different assumed initial
eccentricities.

For small initial amplitudes ($e_{44}=0.01$), this mode must be
excited by the secular resonances to values near the observed
amplitude.  Resonant excitation from migration parameters $< 0.2$ Myr
is insufficient to perturb the $j=4$ mode to the   observed state (see
the dashed black line in Figure~\ref{fig:e33e44}(b)).  For migration
parameters in the range of 0.5--1.0 Myr, about 50\% of cases yield
values consistent the observed $e_{44}$ amplitude.  As the migration
parameter increases above 1.0 Myr, the percent of cases consistent with
the observed value of $e_{44}$ decreases to about 5\% at about 20 Myr.   

When the initial eccentricity amplitude is larger
(e.g.,~$e_{44}=0.05$), excitation from migration parameters
$\hat{\tau}< 0.4$ Myr is insufficient to alter the initial amplitude
out of the acceptable range in more than 50\% of cases (see the black
dotted line in Figure~\ref{fig:e33e44}(b)).  Even with a large initial
amplitude $e_{44}=0.10$, cancellation between the multiple
contributions produces final amplitudes consistent with the observed
one in $\gtrsim20$\% of cases for a large range of migration
parameters ($\hat{\tau}\simeq 0.1\mbox{--}2$ Myr, see the gray solid line in
Figure~\ref{fig:e33e44}(b)).   

The observed $e_{44}$ amplitude is sufficiently large that it can be
achieved from initial amplitudes in the range of 0.01--0.1 and
migration parameters 0.4--2 Myr with a typical occurrence frequency of
20\%--50\% (or higher depending on specific parameter values and
amplitudes).

\subsection{Implications}

The small observed $e_{33}$ amplitude provides a stronger constraint
on giant planet migration than the $e_{44}$ amplitude.  Specifically,
the primordial, pre-migration $C_3$ mode amplitude must be less than
or  comparable to its current value (i.e.,~$e_{33}\lesssim0.01$) and
the migration parameter $\hat{\tau}\lesssim2$ Myr to account for $j=3$
amplitude without requiring the low probability cancellation between
large contributions.  

If we consider the evolution of the $j=3$ and $j=4$ modes as
independent, we can combine the analysis of the individual modes to
ascertain the probability of these resonance producing a terrestrial
system with \emph{both} $e_{33}$ and $e_{44}$ close to their observed
values as the product of the probability of two independent events.
For initial amplitudes  of $e_{33}\lesssim0.01$ and $e_{44} \lesssim
0.05$ there is a reasonable range of migration parameters (0.6--1 Myr)
where the constraints on both $e_{33}$ and $e_{44}$ can individually
be satisfied in $\gtrsim50$\% of cases.   While our analysis is
approximate, it suggests that crossing the 2:1 MMR might produce a
state similar to the observed eccentricity partitioning of the $j=3$
and $j=4$ terrestrial modes in $\simeq25$\% of cases with these
initial amplitudes and migration parameters.

\section{The $g_2-\nu_{5}$ and $g_1-\nu_5$ Secular Resonances}
\label{sec:g21}

For eccentricities of Jupiter and Saturn comparable to their current
values as they passed through the $g_2-\nu_5$ and $g_1-\nu_5$
resonances, our secular theory indicates migration parameters
$\hat{\tau}\gtrsim$0.05 Myr and 0.7 Myr will excite the terrestrial $j=2$
and $j=1$ modes to values larger than the observed ones.  The order of
magnitude difference in the critical parameters between these modes is
primarily due to the larger width ($F_{25}$) of the $g_2-\nu_5$
resonance and the smaller amplitude of the $j=2$ mode.  A migration
timescale of 0.05 Myr is actually \emph{less} than the period of the
resonant frequency ($\approx0.17$ Myr).  For such rapid evolution it is
no longer appropriate to consider the secular mode amplitudes
$\{C_j\}$ as effectively constant during migration.   

We again explore the dynamics of passage through these resonances by
conducting $N$-body simulations of forced giant planet migration with
timescales that span the adiabatic secular ($\tau\simeq1$ Myr) and
non-adiabatic migration ($\tau\simeq0.01$ Myr) regimes.  Migration
timescales as short as $\tau\simeq0.01$ Myr are not likely to result
from smooth migration driven by the scattering of small bodies by
Jupiter and Saturn.  Such rapid migration might result from scattering
with other planets or ice giants.  In this sense our use of smooth
migration with very short migration timescales may serve as a simple
proxy for the rapid non-adiabatic evolution driven by giant planet
scattering.

\subsection{Initial Conditions}

We again isolate the influence of secular resonances by constructing
initial conditions where the terrestrial mode amplitudes are very
small, the giant planets' modal amplitudes are near their present
values,  Jupiter's semimajor axis is fixed at 5.2 AU, and Saturn's
initial semimajor axis of 8.5 AU places the $\nu_5$ frequency between
the $g_3$ and $g_2$ frequencies.  We prepared this initial condition
using a series of orbital integrations.  First, we performed an
integration  starting with only Jupiter and Saturn with their observed
orbits and slowly forced Saturn to migrate inward to a semimajor axis
of 8.5AU.  Because the forced migration is slow relative to the
secular timescale, the $D_{5,6}$ modal amplitudes are effectively
constant during migration to the new orbits.  The terrestrial planets
were then added to the system and we conducted a second integration
applying slow eccentricity damping to decrease the terrestrial mode
amplitudes.  The resulting state of the Jupiter-Saturn-terrestrial
planet system was then used as the initial condition in the numerical
experiments that  follow.  FMFT analysis of this initial condition
finds the terrestrial eccentricity amplitudes with
$e_{ij}\lesssim0.005$ and the $e_{55}$ amplitude of Jupiter very near
its current value (i.e.,~$e_{55}\simeq0.0435$).  Thus, for this set of
simulations $\hat{\tau}=\tau$.

\subsection{Results}

\begin{figure}
\begin{center}

\subfigure{
\resizebox{0.40\textwidth}{!}{\includegraphics{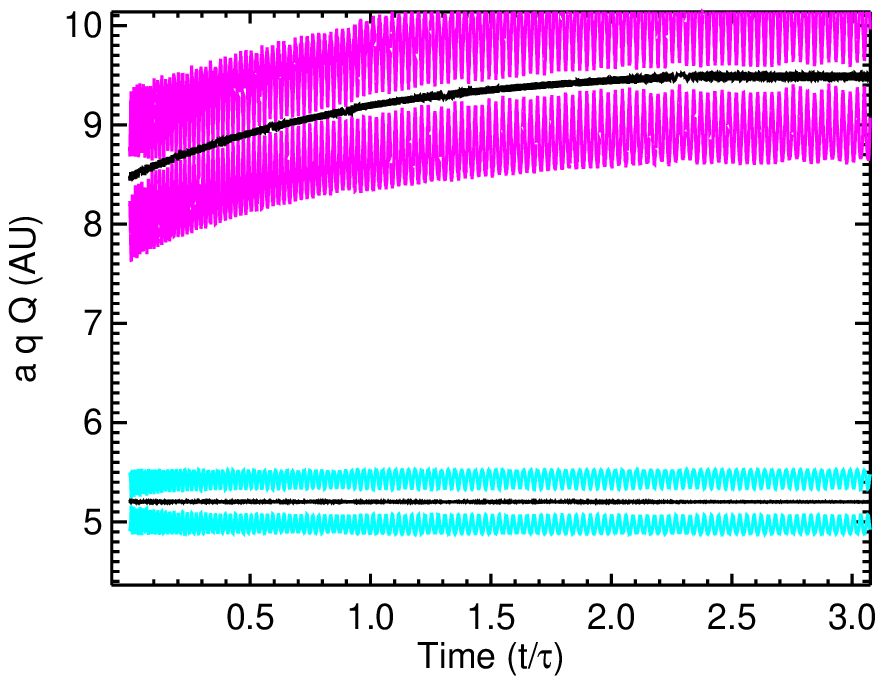}}}\\
\subfigure{
\resizebox{0.40\textwidth}{!}{\includegraphics{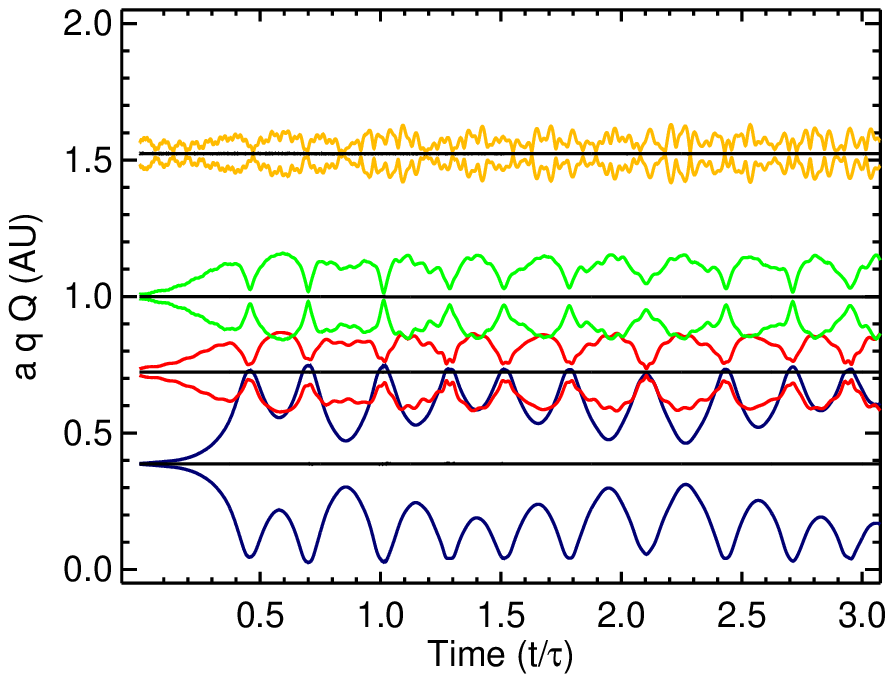}}}

\end{center}
\caption{Forced migration simulation with $\tau = 2.0$ Myr.  The
  semimajor axes of the Jupiter and Saturn and the terrestrial
  planets are shown with a black solid line.  The pericenter and
  apocenter are shown in gray (colored in online version). 
  \label{fig:jsa21_tau1e6}}  
\end{figure}

Figure \ref{fig:jsa21_tau1e6} shows the eccentricity evolution of the
terrestrial planets as Saturn migrates from 8.5AU $\rightarrow$ 9.5 AU
with an exponential timescale of $\tau=2$ Myr.  The $g_2-\nu_5$ and
$g_1-\nu_5$ resonances are crossed at about $t/\tau=$0.3 and 0.6 when
Saturn is near 8.7 and 8.9 AU, respectively.  Excitation of both
terrestrial eigenmodes is evident in the relative timing of the
maximum and minimum eccentricity observed in the orbits of Mercury,
Venus, and Earth.  Mercury's osculating eccentricity reaches a maximum
when contributions from both the $j=1$ and $j=2$ modes are in phase.
For this same orientation between modes, the eigenvector signatures
for Earth and Venus are anti-aligned (see Figure  \ref{fig:e_vects})
resulting in cancellation between modal contributions and the
osculating eccentricities of Earth and Venus simultaneously reach a
minima.  For this cancellation to yield eccentricities near zero, the
eccentricity amplitudes from each mode must be comparable.  In
agreement with secular theory, the eccentricity amplitudes of Mercury,
Venus, and Earth are each driven to values in excess of the present-day values.  Mars is only weakly coupled to the other terrestrial
planets through the $j=1,2$ modes (see Figure  \ref{fig:e_vects}) and
is minimally excited via these resonances. 

\begin{figure}[!h]
\begin{center}
\subfigure[Mercury]{
\resizebox{0.40\textwidth}{!}{\includegraphics{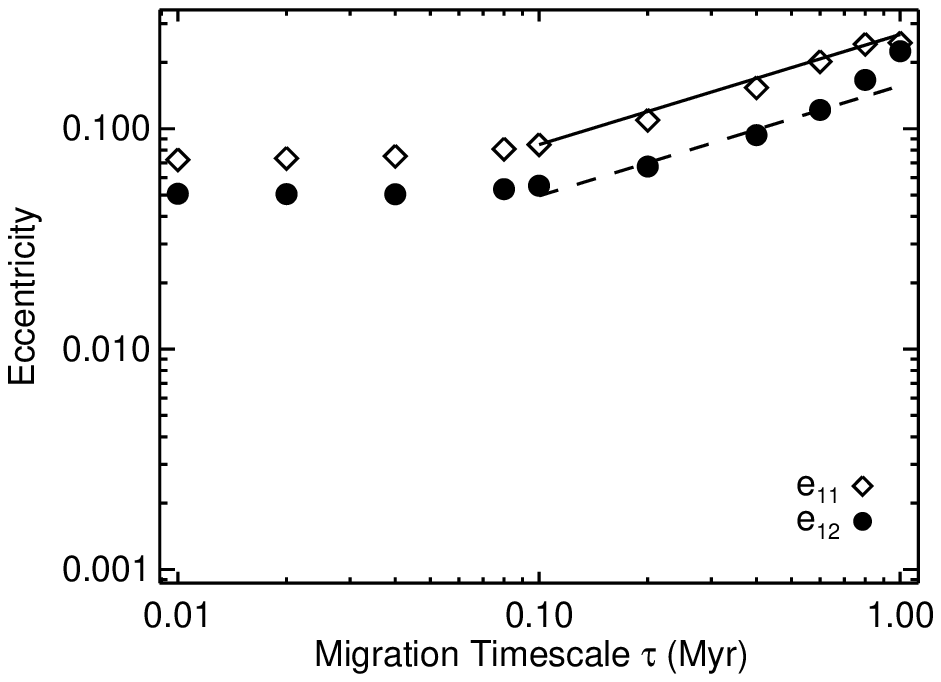}}}
\subfigure[Venus]{
\resizebox{0.40\textwidth}{!}{\includegraphics{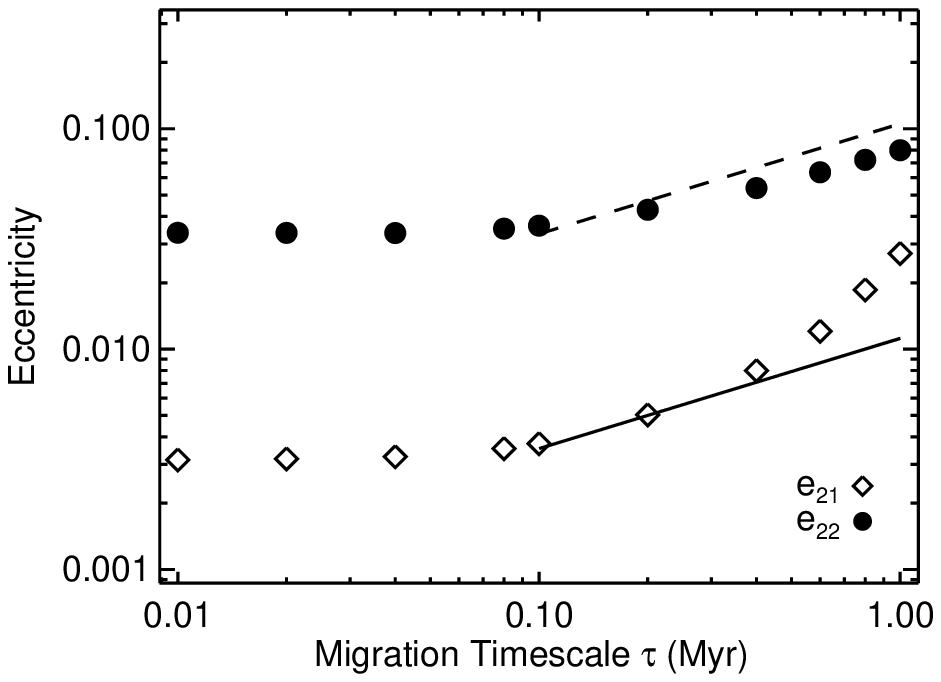}}}
\subfigure[Earth]{
\resizebox{0.40\textwidth}{!}{\includegraphics{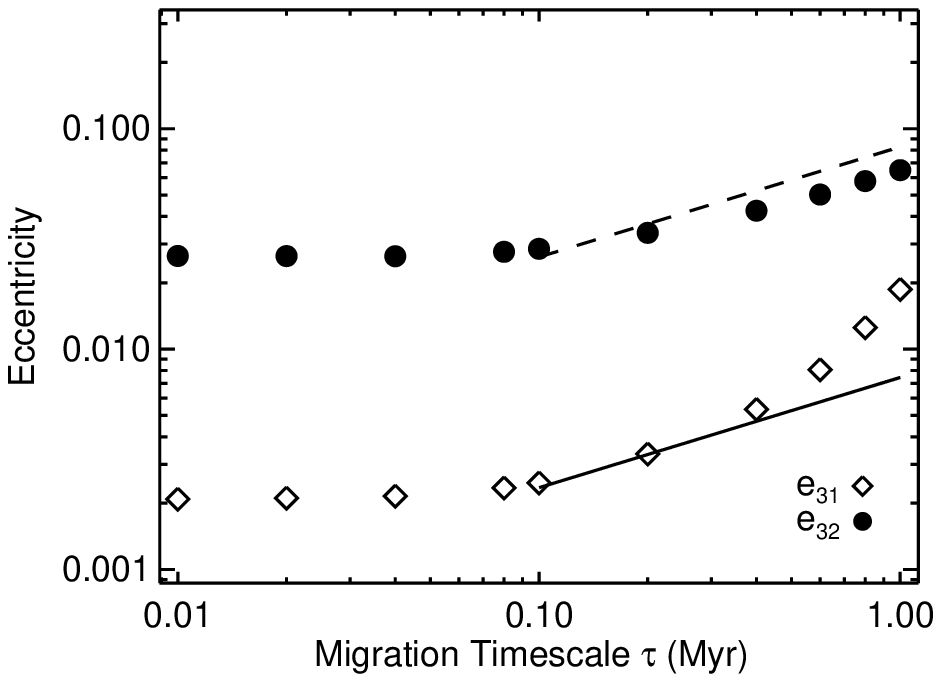}}}

\end{center}
\caption{Final modal eccentricity amplitudes ($e_{ij}$) of the
  terrestrial planets as a function of the giant planet migration
  timescale ($\tau$).  The symbols indicate
  FMFT measurements of the amplitude from $N$-body simulations and the lines
  indicate the prediction of secular theory for the excitation of the
  $j=1,\,2$ modes in each planet (Equation~(\ref{eq:erscale})). Results for the
  $j=1$ mode are shown with solid lines and open diamonds and 
  results for $j=2$ mode are shown
  with dashed lines and filled circles. \label{fig:dekj_g21}}      
\end{figure}

Following each migration simulation we conducted a separate orbital
integration and performed FMFT analysis of the resulting dynamical
state.  The modal eccentricity amplitudes ($e_{ij}$) for Mercury,
Venus, and Earth are shown in Figure \ref{fig:dekj_g21} as a function
of the migration timescale ($\tau$).  The predicted excitation of the
$j=1,\,2$ modes are shown as solid and dashed lines, respectively.  Also
the FMFT-measured eccentricity amplitudes for the $j=1,\,2$ modes are
shown with diamonds and filled circles, respectively.

For migration timescales $\tau=0.1\rightarrow0.4$ Myr, the eccentricity
amplitudes closely match the predictions of secular theory and follow
the $\tau^{1/2}$ scaling.  For Mercury the excitation of
$e_{11}\simeq0.20$ requires $\tau=0.6$ Myr.  Note the $j=2$ component
of Mercury's eccentricity is excited to half that of the $j=1$ mode
($e_{12}\approx0.5e_{11}$) for any timescale $\tau$.  For slower
migration, Mercury may be excited to very large eccentricity due to
the combined effect of both resonances.  In these cases general
relativity and higher order effects conspire to increase Mercury's
precession rate and the value  of the $g_1$ eigenfrequency
\citep{Laskar_2008}.  For  $\tau=0.3\rightarrow1.0$ Myr, the FMFT
measurement of the $g_1$ eigenfrequency increases from
$5.84\arcsec\mbox{yr}^{-1}\rightarrow6.3\arcsec\mbox{yr}^{-1}$.   This tuning
of the $g_1$ toward the $g_2=7.43\arcsec$/yr alters the structure of
both eigenvectors.  As two eigenfrequencies converge, their
eigenvector signatures become increasingly similar \citep[see detailed
discussions of this issue in][]{Ward_1981}.  As the value of $g_1$
increases toward the $g_2$, the magnitude of the components of Earth
and Venus in the $j=1$ eigenvector increase  and those of Mercury
slightly decrease.  The $j=2$ eigenvector evolves in a complementary
fashion.  The component of Mercury in the $j=2$ mode ($v_{12}$)
increases while the components of Venus and Earth ($v_{22}$ and
$v_{32}$) decrease.  In this case, the modest discrepancies in
FMFT-measured eccentricity amplitude and secular theory in Figure
\ref{fig:dekj_g21} result from changes in the eigenfrequencies and
eigenvector signatures.   We note that these effects become important
at eccentricities well above the observed values in the present solar
system.    

For migration timescales $\tau<0.1$ Myr, the entire epoch of migration
may occur in less than a precession cycle.  Such migration is not slow
with respect to the $\nu_5$ frequency.  Since the eccentricity
amplitudes of the terrestrial planets  were all $e_{ij}\lesssim0.005$
prior to migration, Figure \ref{fig:dekj_g21} shows that rapid
non-adiabatic giant planet migration can also contribute to the
eccentricity amplitudes and AMD in the terrestrial modes. For the
$j=2$ mode, the eccentricity amplitudes of $e_{12}=0.052$,
$e_{22}=0.033$, and $e_{32}=0.025$ for Mercury, Venus, and Earth are
systematically achieved and appear independent of the migration
timescale.  These values correlate with the planet's distance from the
$\nu_5$ resonance in the present-day solar system  and are likely
related to the forced   eccentricity in the region.  These values are
also \emph{greater} than the observed values for the $j=2$
eccentricity amplitudes of the terrestrial planets.  

For the fastest migration timescale explored above, this change in
eccentricity amplitudes results from a single rapid jump from one
orbital state of Jupiter and Saturn to another.  However, giant planet
migration due to scattering between the planets would involve several
to many stochastic jumps in semimajor axes and modal amplitudes
\citep{Morbidelli_etal_2009}.  Repeated scattering events among the
giant planets may induce stochastic and diffusive exchange of AMD
between the terrestrial and giant planet mode amplitudes.  If the
eccentricities shown in Figure \ref{fig:dekj_g21} are indicative of
the size of the steps, then this type of evolution may also perturb
the terrestrial system to eccentricities larger than the observed
values.  As in the case of slower planetesimal-driven migration,
selective cancellation between multiple contributions may be required
for the terrestrial planets to emerge from this type of giant planet
evolution with an eccentricity state resembling the observed one, even
if the initial $j=2,\,3$ mode amplitudes are small.
\citet{Brasser_etal_2009} show an $N$-body simulation result that
illustrates this type of evolution.   A systematic exploration of the
diffusion of AMD between the giant and terrestrial planets during an
epoch of giant planet scattering is needed to more generally evaluate
the influence this style of giant planet migration has on the
terrestrial planets.

\subsection{Interference of Resonant and Primordial Amplitudes}

There is about an order of magnitude difference between the critical
migration parameters for the $j=2 \mbox{ and } 1$ modes
($\hat{\tau}^*_{25}= 0.05 $ Myr and $\hat{\tau}^*_{15}= 0.7$ Myr, see
Table \ref{tab:secres}) with both values shorter than migration
timescales suggested for planetesimal-driven migration by at least an
order of magnitude.  Could cancellation between the resonant
contributions yield a dynamical state consistent with the terrestrial
planets for migration timescales of 2--10 Myr?   To evaluate this
hypothesis we model the addition of multiple contributions to the
$j=2,\,1$ modes using the Monte Carlo technique described in Section
\ref{sec:interfere}.  Because the $g_2-\nu_5$ and $g_1-\nu_5$
resonances are crossed once during the last $\sim1$ AU of divergence
between Jupiter and Saturn, the final mode amplitude results from just
two contributions, the secular resonant one and the primordial
amplitude.   When comparing results with the terrestrial planets, we
consider ``close'' to the observed dynamical state to mean $e_{22}<0.03$
and $0.12<e_{11}<0.23$ for the $j=2$ and $j=1$ modes, respectively.  As
above, these ranges are somewhat arbitrary, but are comparable to the
observed $e_{22}=0.02$  amplitude and within 30\% of the observed
$e_{11}=0.18$ amplitude.  

\begin{figure}[ht!]
\begin{center}
\subfigure[Fraction with $e_{22}<0.03$]{
\resizebox{0.45\textwidth}{!}{\includegraphics{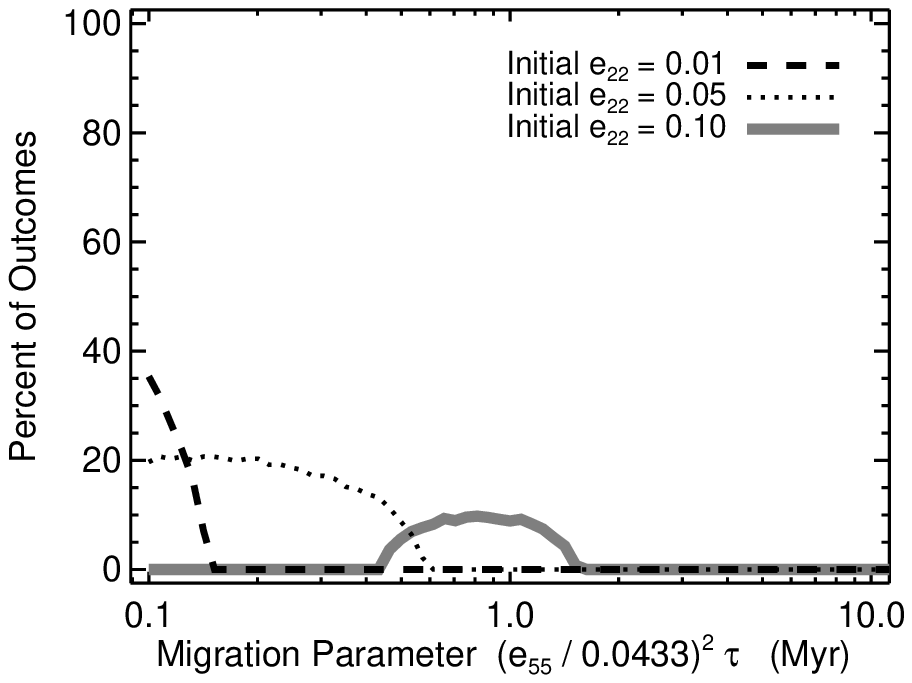}}}
\subfigure[Fraction with $0.12 < e_{11}< 0.23 $]{
\resizebox{0.45\textwidth}{!}{\includegraphics{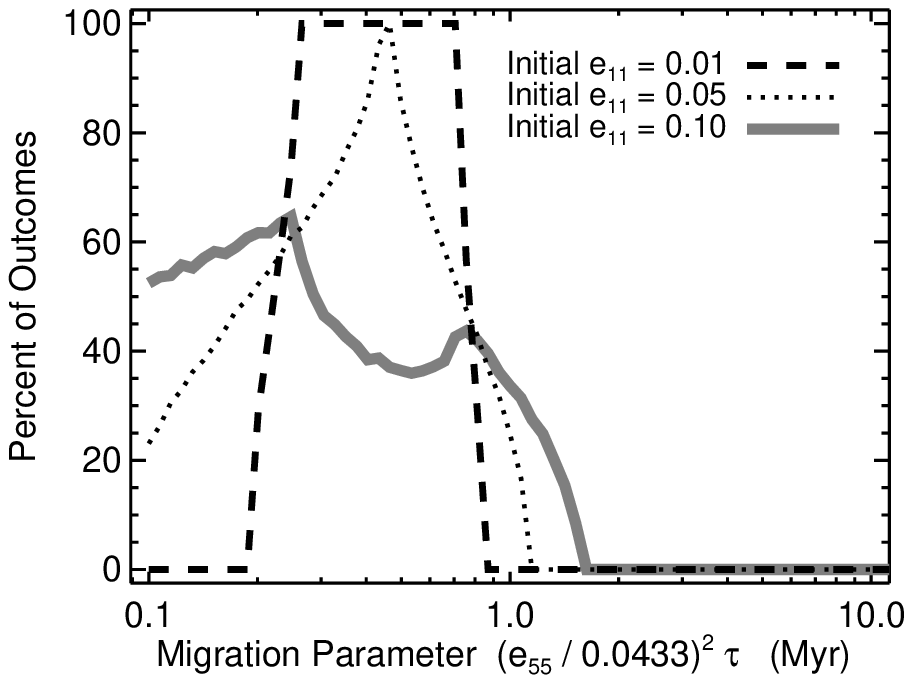}}}
\end{center}

\caption{Percent of outcomes where the $e_{22}$ and $e_{11}$
  eccentricity amplitudes are close to the observed dynamical
  state of the terrestrial planets are shown as a function of the
  migration parameter (Equation~(\ref{eq:mp})).  Contributions from the
  $g_2-\nu_5$ and  $g_1-\nu_5$ resonances are combined
  with an assumed initial amplitude as in Figure
  \ref{fig:e33e44}. Curves for different assumed initial
  eccentricities are shown. 
   \label{fig:e22e11}}  
\end{figure}

Figure \ref{fig:e22e11} shows the percent of outcomes consistent with
the observed $e_{22}$ and $e_{11}$ as a function of the migration
parameter for several different assumed primordial eccentricity
amplitudes.  Cancellation can produce systems with $e_{22}<0.03$ for
migration parameters larger than the critical one (see, e.g.,~Figure
\ref{fig:e22e11}(a)), but the requisite destructive interference
requires primordial and resonant contributions of comparable size.
Also, the probability of the required cancellation decreases as the
size of contributions (and migration parameter) increases.  For
example, migration parameters of 1 Myr contribute a vector of about
0.10 in magnitude to $e_{22}$ (see Equation~(\ref{eq:erscale} and Table
\ref{tab:secres}).  Combining this with an initial eccentricity
amplitude of 0.10 will produce a sufficiently small final value of
$e_{22}$ in $\lesssim10$\% of cases.    

Figure \ref{fig:e22e11}(b) shows results for the $e_{11}$ amplitude.
Cancellation with an initial $e_{11}$ amplitude generally broadens the
range of migration parameters that are able to produce systems close
to the observed state of the terrestrial planets.  However, even large
initial eccentricities of 0.10 do not permit migration parameters much
larger than 1 Myr, and are not consistent with migration timescales
suggested for planetesimal migration, unless the amplitude of the
$e_{55}$ amplitude is lower than its observed value as these
resonances are encountered.

Recall that the $N$-body simulations with migration timescales
$\tau=0.01$ Myr are fast with respect to the secular timescale.  Also
for very fast migration the resulting $e_{11}$ and $e_{22}$ amplitudes
are comparable to those obtained by migration parameters
$\hat{\tau}\simeq 0.1$ Myr.  We use this level of excitation to
estimate the minimum forcing expected during instability-driven
migration (i.e.,~planet-planet scattering).   For example, Figure
\ref{fig:e22e11} shows that for modest initial eccentricity amplitudes
(i.e.,~$e_{11}\simeq e_{22}\simeq 0.05$) and fortuitous interference
between primordial and resonant contributions, both the large observed
value of $e_{11}$ and the small observed value of $e_{22}$ might be
produced in $\sim10$\% of cases for migration parameters of
$0.10\mbox{--}0.4$ Myr.  This fast exponential migration modeled here might be
thought of as the rapid divergence of Jupiter and Saturn by a single
strong scattering event, or one large jump in semimajor axis to the
observed orbit.  When  Jupiter and Saturn experience multiple
scattering events the forcing of the terrestrial planets may be
greater and the success rate in producing the observed state of the
inner planets lower.

Because the $e_{11}$ modal amplitude  is about an order of magnitude
larger than $e_{22}$ and both are excited in comparable ways by the
sweeping $\nu_5$ resonance (see Equation~(\ref{eq:erscale}), Table
\ref{tab:secres}, and Figure \ref{fig:dekj_g21}), it is difficult to
explain their very different amplitudes as a natural outcome of
planetary migration, without invoking strong interference between
multiple  contributions or additional processes.

\section{Discussion}
\label{sec:disc}

Using a modified secular theory we have calculated how the $\nu_5$ and
$\nu_6$ eigenfrequencies of the Jupiter--Saturn system evolve as these
planets diverge from just outside their mutual 3:2 mean motion
resonance to their observed orbits.  Using the subgroup approximation
of \citet{Ward_1981} we have identified six secular resonances between
the $\nu_5$ frequency and one of the terrestrial eigenfrequencies
$g_{j=1\mbox{--}4}$ during this divergent migration.  

Based on analytic arguments we combined the Jovian eccentricity
amplitude ($e_{55}$) and orbital migration timescale ($\tau$) into a
single migration parameter $\hat{\tau}=(e_{55}/0.0433)^2\tau$ that
encapsulates the forcing strength experienced by the terrestrial
planets via secular resonances with the $\nu_5$ frequency.  We used
this model to identify the critical values of the migration parameter
required to excite the terrestrial eccentricity eigenmodes to their
observed amplitudes.   We validated the analytic secular model with
$N$-body simulations and showed that linear secular models can account
for the fundamental character and scaling of the eccentricity
excitation experienced by the terrestrial planets via these secular
resonances.  We then used this model and the observed modal amplitudes
of the terrestrial planets to constrain the migration of Jupiter and
Saturn when these secular resonances were encountered.

\subsection{Evaluating Model Assumptions} 

To clarify the dominant processes at work, we have made a number of
simplifying assumptions.  First, we have neglected the influence of
Uranus and Neptune.  The distant locations and relatively small mass
of the ice giants render them weak perturbers of Jupiter and Saturn.
Their omission from the secular  model does not significantly modify
the $\nu_5$ and $\nu_6$ frequencies of Jupiter and Saturn or those of
the terrestrial planet subsystem (e.g.,~$\Delta
g_1/g_1\sim\mathcal{O}(10^{-3})$).  

Similarly, we have neglected any modification of the system's
eigenfrequencies that results from interaction with the planetesimal
disk fueling migration.  The modification of the $\nu_5$ and $\nu_6$
frequencies by the planetesimal disk is a function of the disk mass
and location.   When considering the late migration of Jupiter and
Saturn, the planetesimal disk must have survived for at least the
30--100Myr required to form the terrestrial planets.   Bodies with
orbits near Jupiter and Saturn have a median dynamical lifetime of
about 6--9 Myr
\citep{Tiscareno_&_Malhotra_2003,Bailey_&_Malhotra_2009}. This
suggests that the Jupiter--Saturn region was largely devoid of
planetesimals by the time the terrestrial planets were completely
assembled \citep[see, e.g.,][]{Gomes_etal_2005}.  A planetesimal disk
exterior to Jupiter and Saturn with mass comparable to that of Uranus
and Neptune (i.e.,~$\sim30M_{\oplus}$)  affects the $\nu_5$ and $\nu_6$
frequencies by an amount comparable to that of the ice giant planets
and can be similarly neglected.   

For rapid migration ($\tau< 1\mbox{--}2$ Myr), a large fraction of the
planetesimal disk must be on crossing orbits with Jupiter and Saturn.
To roughly estimate the effect of this disk on the giant planet
eigenfrequencies we treat it as a single $10M_{\oplus}$ planet between
the gas giants.  This increases both the $\nu_5$ and $\nu_6$
frequencies by $\simeq20$\%.   However, this contribution  decays as
migration proceeds and the planetesimal disk is depleted.  We suggest
that these deviations in the $\nu_5$ frequency are not likely to
prevent resonances with the terrestrial modes or grossly modify the
factors that determine the excitation experienced by the inner
planets.   

In exploring the problem we held Jupiter's semimajor axis constant to
reduce the number of model parameters.  More generally, the
configuration of the planetary system when secular resonances occur
depends on the semimajor axes of all the planets. So, the examples
shown and values listed in Table \ref{tab:secres} are not unique.
Self-consistent $N$-body simulations of planetesimal-driven migration of
the solar system typically show Jupiter to migrate inward a few tenths
of an AU and Saturn migrating outward a few AU on the same timescale
\citep[e.g.,][]{Hahn_&_Malhotra_1999,Gomes_etal_2004}. 

We have used the secular analytic model to examine how Jupiter's
migration affects the planetary configuration where resonances  occur
and the critical migration parameter for each resonance.   Considering
Jupiter with an initial semimajor axis of 5.4 AU, Saturn starting just
outside the 3:2 MMR with Jupiter, and both planets migrating to their
observed orbits,  all six of the secular resonances reported in Table
\ref{tab:secres} are encountered, albeit with slightly different
values of the planetary semimajor axes, resonant frequencies, and
critical migration timescales.  While the $\nu_{5,6}$ frequencies are
strongly affected by the orbital divergence of Jupiter and Saturn, the
effect of Jupiter's inward migration of $\sim$0.2 AU on the terrestrial
eigenfrequencies ($g_{1\mbox{--}4}$) is small and only slightly larger than
the effect of Saturn's outward migration (see,
e.g.,~Figure~\ref{fig:gjvsas}).   Consequently, the orbital period ratio
between Saturn and Jupiter ($P_S/P_J$) at a particular $g_j-\nu_5$
secular resonances is nearly independent of Jupiter's semimajor axis,
or whether it is migrating.  This allows the period ratio to be used
as a single independent variable to identify secular resonances
\citep[see, e.g.,][]{Brasser_etal_2009} and for comparison with
scenarios where both gas giants are migrating.  

Further, the secular model shows that the frequency gradient at
resonance due to Jupiter ($|d(g_j-\nu_5)/da_5|$) is comparable to that
of Saturn ($|d(g_j-\nu_5)/da_6|$).  In this situation, the planet with
the greater migration rate (i.e.,~$da/dt$) most strongly determines the
level of terrestrial excitation and the critical migration parameter
(see Equation~(\ref{eq:tau_mode})).   Due to its larger mass Jupiter tends to
be less mobile and its planetesimal-driven migration rate generically
slower than Saturn's.  Consequently, Saturn's migration rate most
strongly determines excitation of the inner solar system.   Finally,
for the cases where Jupiter migrates inward a few tenths of an AU,
the critical migration  parameters vary from those listed in Table
\ref{tab:secres} by 20\%--30\%. 

In total, the various assumptions we have made to simplify the model and
its presentation modestly affect the secular system.  We consider the
critical migration parameters listed in Table \ref{tab:secres} as
broadly  representative of many plausible migration scenarios
suggested for the solar system.

\subsection{Constraints on the Migration of Jupiter and Saturn}

As Jupiter and Saturn diverge across their mutual 2:1 MMR they
encounter the $g_3-\nu_5$ and $g_4-\nu_5$ secular resonances on both
sides of the resonance (i.e., for $P_S/P_J\simeq1.97\mbox{--}2.03$).  Due to
several factors, including the small amplitude of the $j=3$
terrestrial mode, the $g_3-\nu_5$ resonances near the 2:1 MMR of
Jupiter and Saturn constrain the giant planet migration parameter to
satisfy $ \hat{\tau} \lesssim  \hat{\tau}_{35}^{*} = 0.60$
Myr. Following the 2:1 crossing, the $g_2-\nu_5$ and $g_1-\nu_5$ 
resonances are encountered as Jupiter and Saturn diverge outside their
2:1 MMR (when $P_S/P_J\simeq$2.1--2.2, see Table \ref{tab:secres}).  In
a similar manner the migration history considered must satisfy
$\hat{\tau} \lesssim  \hat{\tau}_{25}^{*} = 0.05$ Myr when
$P_S/P_J\simeq2.1$ and $\hat{\tau} \lesssim \hat{\tau}_{15}^{*} =
0.68$ Myr when $P_S/P_J\simeq2.2$.  We note that greater
values of the migration parameters may also produce acceptable states
for secular resonances with the terrestrial modes ($j=1\mbox{--}4$).  However,
achieving agreement with the observed terrestrial system relies on
cancellation of multiple contributions (see,
e.g.,~Figures~\ref{fig:interfere} and \ref{fig:dekj_js21}).   When the
magnitudes of all contributions are known, we have shown that the
probability of producing a particular eccentricity state can be
estimated using analytical arguments (see~Equation~(\ref{eq:ef}),
Figures~\ref{fig:e33e44} and \ref{fig:e22e11}).  We have also shown that
the probability of yielding the required cancellation between multiple
contributions decreases significantly when the critical migration
parameter $\hat{\tau}_{j5}^*$ is exceeded by a factor of two to three.  With
these caveats we suggest that these values represent characteristic
constraints on the migration of Jupiter and Saturn due to passage
through these secular resonances.   Further, because the solar
system's giant planets migrate as a coupled system, constraints on the
migration of Jupiter and Saturn should be considered as indirect
constraints on the formation and migration of Uranus and Neptune and
the evolution of the giant planet system as a whole.  

For each $g_j-\nu_5$ secular resonance identified, planetesimal-driven
migration of the giant planets (i.e.,~$\tau\simeq 5\mbox{--}20$ Myr) with the
Jovian eccentricity comparable to the present value would excite the
terrestrial system to eccentricities and modal amplitudes well beyond
their observed values.  The constraints on giant planet migration
imposed by these secular resonances and the dynamical structure of the
terrestrial planets may be satisfied in several ways.  

First, if the giant planets migrated with eccentricity amplitudes
comparable to their present day values the strong forcing of the
terrestrial planets via the secular resonances may have interfered
destructively with an initial amplitude of the $j=1\mbox{--}4$ modes
(e.g.,~from the accretion epoch). This could deposit the terrestrial
planets in a dynamical state similar to the observed one.  The general
scenario requires the cancellation of two large components and we have
shown that this is accordingly of low probability \citep[see
also][]{Brasser_etal_2009}.   

Using the resonant excitation model and geometric arguments we can
identify additional constraints required on this type of evolution.
For cancellation to be viable, the initial and resonant contributions
must be of a similar size and the excitation from resonance crossing
necessarily less than an amount that would  render the terrestrial
subsystem unstable.  For the $g_3-\nu_5$ and $g_4-\nu_5$  secular
resonances encountered near the 2:1 MMR, the terrestrial planets are
driven to crossing orbits if the migration timescale of Jupiter and
Saturn is $\tau\gtrsim10$Myr and the $e_{55}$ has the present-day
value.  For the $g_2-\nu_5$ and $g_1-\nu_5$ resonances, $\hat{\tau}
\gtrsim 2$Myr leads to crossing orbits and instabilities (see,
e.g.,~Figure~\ref{fig:jsa21_tau1e6}).  This latter constraint indicates
that planetesimal-driven migration through the $g_{2,1}-\nu_5$ secular
resonances is not viable unless the Jovian eccentricity amplitude
$e_{55}$ is significantly smaller than the observed value.

Second, if the Jovian eccentricity ($e_{55}$) was smaller during
migration, then longer timescales might be permissible.  For
$\tau\sim10$ Myr, the $g_3-\nu_5$ secular resonance constrains the
Jovian eccentricity amplitude $e_{55}\lesssim0.011$, or about a
quarter of the current value, when Jupiter and Saturn migrated across
the 2:1 MMR.  This eccentricity is comparable to those typical of
models of planetesimal-driven migration.   However, since it is less
than the observed $e_{55}=0.0433$ value it also requires that the
$D_5$ mode amplitude be excited later via some other
process. \citet{Morbidelli_etal_2009} examined the origin of the $D_5$
mode amplitude in detail and found that repeated planet--planet
scattering between Jupiter or Saturn and an ice giant planet may
explain the observed $D_5$ amplitude and account for its excitation
via an instability in the giant planets' orbits after the 2:1 MMR of
Jupiter and Saturn is crossed.  This analytic result is consistent
with the numerical simulations of \citet{Brasser_etal_2009} and the
conclusions drawn from them.  Using the secular model we have also
shown for initial eccentricity amplitudes $e_{33}\lesssim0.01$ and
$e_{44}\lesssim0.05$ the eccentricity partitioning of the $j=3$ and
$j=4$ terrestrial modes can be expected with a roughly 25\%
probability for migration parameters in the range
$\hat{\tau}=$0.6--1.0 Myr.  Again, these values are broadly consistent
with those of planetesimal-driven migration.

When examining the $g_2-\nu_5$ and  $g_1-\nu_5$ resonances, a Jovian
modal eccentricity amplitude of $e_{55}\lesssim 0.003$ is required to
allow a planetesimal-driven migration timescale of $\tau\simeq10$Myr
through these resonances without exciting the $j=2$ terrestrial mode
amplitude beyond its observed value.  This value of $e_{55}$ is quite
small, and again requires the late excitation of the Jovian
eccentricity \citep[e.g.,~via scattering with a  $\gtrsim10M_{\oplus}$
body to excite the $D_5$ amplitude during the last 0.5--0.8 AU of
divergent migration between Jupiter and Saturn;
][]{Morbidelli_etal_2009}.
 
Third, rapid, stochastic migration in large steps in semimajor axis is
also  possible when the system of planets is unstable and
planet--planet scattering ensues.  Such global instabilities may allow
secular  resonances to be jumped over and the whole migration epoch to
be much  shorter than suggested by planetesimal-driven migration.
\citep{Thommes_etal_1999,Tsiganis_etal_2005,Morbidelli_etal_2007,Batygin_&_Brown_2010}.
If such rapid migration was driven by scattering between planets, this
process must cause the orbits of Jupiter and Saturn to diverge.  For
Saturn to move outward it must scatter an ice giant inward.  Similarly
for Jupiter to move inward, it must scatter an ice giant outward.
This style of instability-driven migration might allow the $\nu_5$ and
$\nu_6$ secular resonances to move quickly through the inner solar
system.  $N$-body simulations of this process by
\citep{Brasser_etal_2009} suggest that it is possible that the
terrestrial planets might emerge from this style of planet migration
with orbits comparable to their observed ones.  

On the other hand, our simulations with very fast migration
($\tau<0.2$ Myr) indicate that the $j=2$ terrestrial mode can be
excited to amplitudes in excess of the observed values.  If the $D_5$
mode amplitude originated via repeated planet--planet scattering
\citep{Morbidelli_etal_2009} it appears plausible that
instability-driven giant planet migration may perturb the terrestrial
planets to an overly excited state.  In this regard, additional work
assessing the relative frequency of characteristic outcomes
(e.g.,~those similar to the observed state) would be welcome. 

Finally, if planetary migration was rapid as suggested by the
constraints of these secular resonances ($\tau\lesssim$5--10 Myr), then
entire epoch of giant planet formation and migration, could have been
largely complete before terrestrial planet accretion was finished.  In
this case the dynamical structure of the inner solar system was
determined later (e.g.,~by processes related to accretion and
evolution) and the formation timescale of the terrestrial planets
(i.e.,~30--100 Myr) constrains the timing by which giant planet
formation and migration was largely complete.   Recently,
\citet{Walsh_&_Morbidelli_2011} have investigated the effect of the
early planetesimal-driven migration of Jupiter and Saturn on
terrestrial planet formation.  Their $N$-body accretion simulations with
a migration timescale of $\tau=5$ Myr for Jupiter and Saturn produce
terrestrial planet systems with AMDs comparable to the observed
terrestrial planets.  While several open problems of terrestrial
planet formation persist in these simulations (e.g.,~explaining the
small mass of Mars and the clearing and dynamical structure of the
asteroid belt), these results suggest that effects of the sweeping
$\nu_5$ resonance through the terrestrial region may be tempered by
terrestrial accretion dynamics.

\subsection{Implications for the LHB}

In the Nice model explanation of the LHB \citep{Gomes_etal_2005}, an
instability among the  giant planets is initiated by Jupiter and
Saturn crossing their mutual 2:1 MMR.  This instability may be delayed
by the very slow divergent migration of Jupiter and Saturn of a few
tenths of an AU toward the 2:1 MMR over 600--700 Myr.  This slow
migration roughly corresponds to exponential migration timescales of
$\tau\gtrsim 3\times10^{9}$yr.  The secular resonant model presented
here predicts that crossing the $g_3-\nu_5$ and $g_4-\nu_5$ resonances
this slowly requires the Jovian eccentricity amplitude $e_{55}
\lesssim \mathcal{O}(10^{-4})$ to leave the terrestrial planets in a
state comparable to the observed one.  A modestly larger Jovian
eccentricity ($e_{55}\gtrsim0.001$) may perturb the terrestrial
planets to crossing orbits and drive a global instability of the
terrestrial system.   

Requiring such small values of the Jovian eccentricity is a very
strong constraint on the evolution of the gas giants and we interpret
this as an indication that the slow divergent migration of Jupiter and
Saturn toward the 2:1 MMR is not generally consistent with the
coexistence and dynamical structure of the terrestrial planets.  As a
result, the specific LHB initiating scenario described in
\citet{Gomes_etal_2005} appears unlikely.  \citet{Brasser_etal_2009}
have drawn similar conclusions based on the results of $N$-body
simulations.  

If the LHB resulted from the large scale migration of the giant
planets, then an instability and giant planet scattering may be
required to avoid strongly perturbing the terrestrial planets via
secular resonances \citep{Brasser_etal_2009}.  Alternate modes of
triggering a late instability (e.g.,~Uranus and Neptune's divergent
migration across a mean motion resonance) are possible and may offer
viable alternative modes of initiating the LHB
\citep{Morbidelli_etal_2007}.    

However, the $j=5$ mode of Jupiter and Saturn has components in
terrestrial planets comparable  to those of $j=2,\,3$ terrestrial modes
(see, e.g.,~Table~\ref{tab:eij}).  If giant planet scattering is
responsible for exciting the $j=5$ mode amplitude to the observed
value, then this excitation may also be communicated to the
terrestrial planets (e.g.,~stochastic diffusion of AMD  between $j=5$
mode and the terrestrial $j=1$---4 modes).  Our simulations of very fast
migration (i.e.,~$\tau\lesssim0.10$ Myr) suggest that excitation of the
terrestrial planets via this process is likely, but more work to
examine the net exchange of AMD between the terrestrial and giant
planet systems is needed to assess the dynamical implications of
instability-driven giant planet migration for the terrestrial
planets.

\subsection{Summary}
\label{sec:sum}

We have shown that if the terrestrial planets witnessed an epoch when
the gross orbital structure of the  giant planets changed, then the
dynamical structure of the inner solar system was likely altered.  In
this scenario, the ultimate dynamical state of the terrestrial planets
is  a product of several processes (e.g.,~accretion dynamics, passage
through secular resonances, AMD diffusion via giant-planet scattering,
and chaotic diffusion of the system's AMD over long timescales) that
act in concert and whose individual influences are challenging to
isolate and disentangle. 

The small amplitudes of the $j=2,\,3$ terrestrial eigenmodes, that are
closely associated with the small eccentricities of Earth and Venus,
provide strong constraints on the migration of Jupiter and Saturn.
While more work examining the characteristic forcing of the
terrestrial planets by instability-driven migration is needed,
explaining the observed dynamical state of the terrestrial planets as
a natural and frequent outcome of the late migration of the giant
planets remains a challenge. 

An alternate and perhaps simpler possibility is that giant planet
formation and migration was largely complete before  the dynamical
structure of the inner solar system was determined.  Formation models
and isotopic evidence suggest that terrestrial planet accretion was
complete in 30--100 Myr. The bulk of giant planet migration could easily
be accommodated in shorter time  intervals via some combination of
tidal interaction with a gas disk, planet--planet scattering, and
planetesimal-driven migration.  Any strong perturbation of the inner
solar system resulting from giant planet migration might then be
tempered through a variety of processes that depend on the state of
the forming  planets and the protoplanetary disk at the time of the
disturbance.  Additional mechanisms that may act at this earlier time
include dynamical friction with a disk of planetesimals, orbital
damping due to interactions with a remnant gas disk, or a combination
of dynamical processes \citep{Nagasawa_etal_2005,Thommes_etal_2008}.

In the sequential accretion scenario, the formation timescale for
giant planet cores ($\sim10M_{\oplus}$) in the outer regions of a
minimum mass nebula appear to be considerably longer than the
observationally inferred depletion timescale of protostellar
disks. Ice giant formation at smaller orbital radii (and shorter
orbital periods) is often invoked to overcome this problem, but
requires the subsequent outward migration of the ice giant planets.
However, it is also possible that disk gaps near Jupiter and Saturn's
orbits may have provided migration barriers which eventually led to
the accumulation of planet-building materials and the emergence of
Uranus and Neptune within a few million years (see Section
\ref{sec:intro}). 

The resonant structure of the Kuiper Belt is often interpreted as a
product of planetesimal-driven
\citep{Malhotra_1995,Ida_etal_2000,Hahn_&_Malhotra_2005} or
instability-driven migration \citep{Levison_etal_2008} of the giant
planets. However, the giant planets' outward migration  may also be
induced by their tidal interaction with a viscously  expanding gaseous
nebula \citep{LinPapaloizou_1986} or by a disk which  is undergoing
photoevaporation.  In these cases, giant planet  migration may be
mostly complete in $\sim$10 Myr and could predate the final assembly of
the terrestrial planets. Such an early migration scenario would avoid
the excitation of the terrestrial planets' eccentricities. 

This contingency requires that the delivery of LHB impactors to the
Moon be achieved after the formation, large-scale migration, and/or
global orbital restructuring of the giant planets.  Within the context
of the Nice model migration scenarios, only explaining the LHB
requires that the bulk of giant migration occurred late.  If giant
planet migration was completed before the dynamical state of the inner
solar system was determined, other aspects of the Nice model remain
viable (e.g.,~the capture of Jupiter's Trojans and the orbital properties
of the giant planets).  Obviously, the delivery of the LHB impactors
must then be accounted for via some other means \citep[e.g.,~the Planet
V hypothesis described in ][]{Chambers_2007}.

Additional progress in unraveling the sequence of events and processes
that produced the observed solar system might be made by examining the
formation and early evolution of the gas and ice giant planets.  We
note that the nature and details of gas and ice giant formation in the
solar system remain poorly understood.   The basic assumptions and
initial conditions of current migration models (e.g.,~that Jupiter and
Saturn formed in a much more compact configuration) may change as the
formation and early history of the giant planets are clarified.

Finally, we note that migration near and through mean motion
resonances is germane to the orbital evolution of planetary systems.
Consequently, the resonant sweeping processes examined here have
applications among extrasolar planetary systems, satellite systems, and
other dynamical characteristics of the solar system.  Analyses of
these applications will be presented elsewhere.   
 
\acknowledgments 
This work has been supported by NASA  (NNG05G1496,
NNX07A-L13G, NNX07AI88G, NNX08AL41G, and NNX08AM84G), and the NSF
(AST-0908807). We thank Carl Murray and Orkan Umurhan for useful
discussions, Tolis Christou for providing the FMFT code used, Ramon
Brasser and Alessandro Morbidelli for comments on an early draft of
this work, and Renu Malhotra for a thorough and helpful review.

\clearpage


\end{document}